\documentclass[iop,apj]{emulateapj}
\usepackage{amssymb}		 
\usepackage{amsmath, amsthm, amssymb,amsfonts}
\usepackage[english]{babel}
\usepackage{epstopdf}
\usepackage{color}
\usepackage{graphicx}
\usepackage{footmisc}

\def\apjs{ApJS}
\def\apj{ApJ}
\def\apjl{ApJL}

\def\araa{ARA\&A}
\def\mnras{MNRAS}

\def\pnas{PNAS}

\def\prl{PhRvL}
\def\prx{PhRvX}
\def\jgr{JGR}
\def\jgra{JGRA}
\def\zhetf{ZhETF}
\def\ssr{SSRv}
\def\frass{FrASS}

\def\grl{GeoRvL}
\def\pop{PhPl}
\def\pof{PhFl}

\def\jcomp{JCoPh}
\def\jpp{JPlPh}

\def\rvmpp{RvMPP}

\usepackage{hyperref}
\definecolor{darkblue}{rgb}{0.0,0.0,0.3}
\hypersetup{colorlinks,breaklinks,
            linkcolor=darkblue,urlcolor=darkblue,
            anchorcolor=darkblue,citecolor=darkblue} 

\newcommand{\njp}{NJP}

\newcommand{\rmd}{{\rm d}}
\newcommand{\rme}{{\rm e}}
\newcommand{\D}[2]{\frac{\rmd #2}{\rmd #1}}
\newcommand{\pD}[2]{\frac{\partial #2}{\partial #1}}
\newcommand\bb[1]{\mbox{\boldmath{$#1$}}}
\newcommand\grad{\bb{\nabla}}
\newcommand\bcdot{\,\bb{\cdot}\,}
\newcommand\btimes{\,\bb{\times}\,}

\newcommand\bs[1]{\boldsymbol{#1}}

\newcommand{\pegpp}{\textsc{Pegasus}\texttt{++} }

\begin{document}

\title{On stochastic heating and its phase-space signatures in low-$\beta$ kinetic turbulence}
\author{S.~S.~Cerri$^1$}
\author{L.~Arzamasskiy$^{1,2}$}
\author{M.~W.~Kunz$^{1,3}$}
\affiliation{$^1$Department of Astrophysical Sciences, Princeton University, 4 Ivy Lane, Princeton, NJ 08544, USA}
\affiliation{$^2$Institute for Advanced Study, 1 Einstein Drive, Princeton, NJ 08540, USA}
\affiliation{$^3$Princeton Plasma Physics Laboratory, PO Box 451, Princeton, NJ 08543, USA}
\email{E-mail of corresponding author: scerri@astro.princeton.edu}

\begin{abstract}
We revisit the theory of stochastic heating of ions and investigate its phase-space signatures in kinetic turbulence of relevance to low-$\beta$ portions of the solar wind. 
In particular, we retain a full scale-dependent approach in our treatment, and we explicitly consider the case in which electric-field fluctuations can be described by a generalized Ohm's law that includes Hall and thermo-electric effects. These two electric-field terms provide the dominant contributions to stochastic ion heating when the ion-Larmor scale is much smaller than the ion skin depth, $\rho_{\rm i}\ll d_{\rm i}$, which is the case at $\beta\ll1$. Employing well-known spectral scaling laws for Alfv\'en-wave (AW) and kinetic-Alfv\'en-wave (KAW) turbulent fluctuations, we obtain scaling relations characterizing the field-perpendicular particle-energization rate and energy diffusion coefficient associated with stochastic heating in these two regimes. Phase-space signatures of ion heating are then investigated using 3D hybrid-kinetic simulations of continuously driven Alfv\'enic turbulence at low $\beta$ (namely, $\beta_{\rm i}=\beta_{\rm e}=0.3$ and $\beta_{\rm i}=\beta_{\rm e}=1/9$). In these simulations, energization of ions parallel to the  magnetic field is sub-dominant compared to its perpendicular counterpart ($Q_{\|,{\rm i}}\ll Q_{\perp,{\rm i}}$), and the fraction of turbulent energy that goes into ion heating is ${\approx}75$\% at $\beta_{\rm i}=0.3$ and ${\approx}40$\% at $\beta_{\rm i}\simeq0.1$. The phase-space signatures of ion energization are consistent with Landau-resonant collisionless damping and a ($\beta$-dependent) combination of ion-cyclotron and stochastic heating. We demonstrate good agreement between our scale-dependent theory and various signatures associated with the stochastic portion of the heating. We discuss briefly the effect of intermittency on stochastic heating and the implications of our work for the interpretation of stochastic heating in solar-wind spacecraft data. 
\end{abstract}

\maketitle

\section{Introduction}\label{sec:intro}

The solar wind is arguably the most well-diagnosed weakly collisional, magnetized plasma, both in terms of the electromagnetic fluctuations it hosts and the thermodynamics of its constituent particles. It therefore serves as an excellent (and, with some effort, directly accessible) laboratory with which one may discriminate between different theories of magnetized turbulence and the various ways in which such turbulence energizes plasma particles. Indeed, a persistent puzzle in solar-wind research is why the temperature of the solar wind evolves non-adiabatically as it expands, and why this heating occurs preferentially in the direction perpendicular to the local magnetic field \citep[e.g.,][]{Marsch1982,MatteiniGRL2007,HellingerJGRA2011,MarucaPRL2011}. While the solution to this puzzle is known to be connected to the pervasive Alfv\'{e}nic turbulence that is now routinely measured by {\em in situ} spacecraft \citep[e.g.,][]{goldstein95,BrunoCarboneLRSP2013,AlexandrovaSSRv2013,ChenAPJS2020,SahraouiRMPP2020}, the relative contributions to this turbulent heating from different wave-particle interactions are debated.

Much of this debate has been centered on the nature of the turbulent fluctuations and their relative energetic importance at various stages during their nonlinear cascade to increasingly finer scales in both configuration and velocity space \citep[e.g.,][]{LeamonJGR1999,HowesJGR2008,SchekochihinAPJS2009,ChandranAPJ2011,CranmerAPJS2014}. Namely, how spatially anisotropic are typical fluctuations at a given scale? What fraction of those fluctuations ultimately attain cyclotron frequencies? Are the fluctuations at Larmor scales of sufficient amplitude to disrupt the particles' otherwise smooth gyro-motion and heat the plasma appreciably? How do the answers to these questions depend on the plasma properties, such as the ratio of thermal and magnetic pressures, $\beta \doteq 8\pi p/B^2$? This is an indirect way of understanding particle energization in the solar wind: guided by observational constraints \citep[e.g.,][]{HorburySSR2012,ChenJPP2016}, one postulates the characteristics of the fluctuations in the turbulent cascade, models the various particle-energization channels available to those fluctuations, and then infers whether these channels are thermodynamically important by comparing the implied heating and any unique features with the data. Such an approach has been used to find evidence for ion-cyclotron-resonant heating in the solar wind via measured correlations between plasma heating, differential flow between ion species, and magnetic-field-biased temperature anisotropy \citep{KasperPRL2013}. Similarly, correlations between the amplitudes of ion-Larmor-scale magnetic fluctuations and enhanced proton and minor-ion temperatures measured in coronal holes and the bulk solar wind have been taken as evidence for the stochastic heating of ions by low-frequency Alfv\'{e}n-wave (AW) and kinetic-Alfv\'{e}n-wave (KAW) fluctuations \citep{Chandran2010,Bourouaine13,Chandran2013,VechAPJ2017,MartinovicAPJ2019,MartinovicAPJS2020}.

A more direct, but more technically challenging, way of distinguishing between different particle energization mechanisms is through their imprint on the  velocity-space structure of the plasma \citep[e.g.,][]{KleinHowesApJL2016,HowesJPP2017,HowesPOP2017,KleinJPP2017,AdkinsSchekochihinJPP2017,ServidioPRL2017,CerriAPJL2018,PezziPOP2018,KawazuraPNAS2019,LiJPP2019}. For example, it is well known that collisionless Landau damping flattens the particle distribution function in the vicinity of ``Landau resonances'', at which a particle's velocity (in a magnetized plasma, the velocity component parallel to the local magnetic-field direction) matches the phase speed of a wave. This flattening is a consequence of the secular transfer of free energy from the electromagnetic waves to the particles, whether it be via parallel electric fields \citep{Landau1946} or parallel gradients in magnetic-field strength \citep{Barnes1966}. Recently, a clear signature of this transfer (in this case, to the electron population) has been found in data taken in the Earth's turbulent magnetosheath \citep{ChenNatCo2019}. This follows on pioneering work by \citet{MarschTuJGR2001} (see also \citealt{HeuerJGRA2007} and \citealt{HeAPJL2015}) showing plateaus in solar-wind particle distribution functions near the Alfv\'{e}n speed, suggesting velocity-space diffusion due to Alfv\'{e}n/ion-cyclotron fluctuations \citep[e.g.,][]{IsenbergSSRv2001,IsenbergVasquezApJ2019}. Similar velocity-space signatures of ion-cyclotron damping, revealed by applying field-particle correlation techniques to hybrid-kinetic simulations, have been discussed by \citet{KleinJPP2020}.

Non-resonant energization mechanisms, such as stochastic heating, also make an imprint on the velocity space. Adopting the theory of \citet{ChandranAPJ2010}, \citet{KleinChandranAPJ2016} showed that the stochastic  heating of ions by moderate-amplitude, Larmor-scale, electric-field fluctuations ultimately flattens the core of their velocity distribution function along the field-perpendicular direction. Such a flat-top distribution has been observed recently by \citet{MartinovicAPJS2020} using data from {\em Parker Solar Probe}. Formulating and testing such velocity-space diagnostics is particularly important in the case of stochastic heating, since it provides an attractive alternative to other (namely, resonant) mechanisms of particle energization whose phase-space signatures have long drawn the attention of the heliophysics community. This becomes particularly true for situations in which the turbulent cascade exhibits strong spatial anisotropy that inhibits the production of high-frequency waves, and/or for values of $\beta\ll{1}$ at which ions are unable to obtain the Landau resonance \citep{Quataert1998,Hollweg1999}. 

Accordingly, the purpose of this paper is to further elucidate the consequences of stochastic ion heating for the organization of phase space and to sharpen certain aspects of how the theory of stochastic heating can be tested using solar-wind data. The paper is written in two parts. First, we extend the work of \citet{ChandranAPJ2010} and \citet{KleinChandranAPJ2016} to make further predictions for the phase-space signatures of stochastic heating and for their dependence on the properties of the plasma ($\beta$, ion-to-electron temperature ratio) and of the turbulence (\S\ref{sec:theory}). Second, we present results from a new hybrid-kinetic simulation of driven, Alfv\'{e}nic turbulence, which we use to test these predictions (\S\ref{sec:simulations}).  We also demonstrate that intermittency, as revealed in the statistics of the electrostatic potential, enhances stochastic heating, with some particles acquiring large amounts of energy in spatially and temporally localized events. A corollary of our analysis is that an oft-employed conversion of measured ion-Larmor-scale magnetic-field fluctuation amplitudes to bulk ion-velocity fluctuations, which are then used in a formula to determine the expected amount of stochastic heating, becomes increasingly inaccurate at low values of $\beta$, precisely where stochastic heating is expected to be most important (\S\ref{sec:interpretation}). For $\beta\ll{1}$, non-inductive components of the electric field -- namely, the Hall effect and the thermo-electric field -- contribute appreciably to the total electrostatic potential with which the particles interact.

Our work follows on that of \citet{ArzamasskiyAPJ2019}. Those authors presented results from hybrid-kinetic simulations of driven, Alfv\'{e}nic turbulence, and employed several novel diagnostics to quantify the roles of Landau and Barnes damping, stochastic heating, and cyclotron heating -- all of which appeared to be in play -- in the energization and differential heating of plasma particles at $\beta\lesssim{1}$. Taken together, this set of simulations and their analyses suggest that stochastic heating plays an important role in modifying both the velocity distribution function of the ions and the cascade of turbulent energy to sub-ion-Larmor scales in low-$\beta$, collisionless plasmas.

\section{Theory of stochastic ion heating in AW/KAW turbulence}\label{sec:theory}

\citet{ChandranAPJ2010} presented a theory for perpendicular ion heating in the solar wind caused by finite-amplitude, low-frequency, AW/KAW fluctuations occurring on scales comparable to the ion-Larmor scale \citep[following on work by][]{ChenPOP2001,JohnsonGRL2001,WhitePOP2002,VoitenkoApJ2004,BourouaineApJL2008}. In this theory, the ions interact stochastically with a time-varying electrostatic potential, break their magnetic moments, and execute a random walk in perpendicular energy. Here, we generalize this theory to account for a spectrum of critically balanced fluctuations whose electrostatic potential satisfies a generalized Ohm's law. We compute the perpendicular heating rate and energy-diffusion coefficient as functions of the perpendicular plasma beta parameter of the ions, $\beta_{\perp\rm i} \doteq 8\pi p_{\perp\rm i}/B^2$, which is the ratio of thermal pressure of the ions perpendicular to the magnetic-field direction, $p_{\perp\rm i} \doteq n T_{\perp\rm i}$ where $n$ is the ion number density, and the magnetic pressure, $B^2/8\pi$; the electron-to-ion temperature ratio, $\tau_\perp \doteq Z_{\rm i} T_{\rm e}/T_{\perp\rm i}$, where $Z_{\rm i}$ is the ion charge in units of $e$; and the energy cascade rate, $\varepsilon$. (We take the electron temperature $T_{\rm e}$ to be isotropic, for reasons that will be explained in \S\ref{subsec:theory_OhmLawArguments}.) Before doing so, we recapitulate briefly the theory presented in \citet{ChandranAPJ2010} in a way that establishes the notation used in the remainder of the paper.

\subsection{Stochastic heating revisited}\label{subsec:theory_revisited}

Consider an ion with mass $m_\mathrm{i}$ and charge $q_\mathrm{i}=Z_\mathrm{i}e$ that is interacting with electric-field fluctuations $\delta\bb{E}_{\perp,\lambda}$ having perpendicular wavelength $\lambda$ of the order of the ion's gyro-radius $\rho_\mathrm{i}\doteq w_\perp/\Omega_\mathrm{i}$, i.e., $k_\perp\rho_\mathrm{i}\sim1$. Here, $w_\perp$ is the component of the ion's random velocity perpendicular to a background magnetic field $\bb{B}_0$, $\Omega_\mathrm{i}\doteq q_\mathrm{i}B_0/m_\mathrm{i}c$ is the ion-cyclotron frequency, and $k_\perp=2\pi/\lambda$ is the field-perpendicular wavenumber associated with $\lambda$. 
If the amplitude of these fluctuations is sufficiently large (just how large is quantified in \S\ref{subsubsec:theory_exp_suppression}), the ion's gyro-motion about $\bb{B}_0$ becomes chaotic, its magnetic moment $\mu\doteq m_\mathrm{i}w_\perp^2/2B$ is no longer conserved, and the ion is stochastically heated in the field-perpendicular direction. 
Such stochasticity is the result of a sequence of ``random kicks'' that the ion experiences due to the fluctuating field within a turbulent eddy of size $\lambda\sim\rho_{\rm i}$. 

In what follows, we assume that the main contribution to this heating is from the potential part of the fluctuating electric field, so that  $\delta\bb{E}_{\perp,\lambda}\sim \bb{k}_\perp \delta\Phi_\lambda$.
This is justified (and verified {\it a posteriori} using our simulations) if  $\beta_{\perp\rm i}$ is not much larger than unity and/or if the fluctuations' frequency $\omega$ remains smaller than ${\sim}\Omega_\mathrm{i}/\beta_{\perp\rm i}$~\citep{Hoppock2018}. Such electrostatic fluctuations on the scale of an ion's gyro-radius induce a change in an ion's perpendicular kinetic energy, $\Delta K_\perp$, that is directly related to the average change of the potential over the time $\tau_\lambda$ that the particle spends within the turbulent eddy of size $\lambda$, {\it viz.}, $\Delta K_\perp\sim q_\mathrm{i}(\overline{\partial\delta\Phi_\lambda/\partial t})\tau_\lambda$. We estimate $\tau_\lambda$ as the time required for the ion's guiding center to drift in the direction perpendicular to $\bb{B}_0$ by a distance of order $\lambda$. Taking this drift to be of the $\bb{B}\btimes\grad\Phi$ type, so that $u_{\rm dr,\lambda}\sim (c/B_0) (|\delta\Phi_\lambda|/\lambda)$, we find that
\begin{equation}\label{eq:taulambda}
    \tau_\lambda \sim \Omega_\mathrm{i}^{-1} \left(\frac{\lambda}{\rho_\mathrm{th,i}}\right)^2 \left(\frac{m_\mathrm{i}v_\mathrm{th,i}^2}{q_\mathrm{i}|\delta\Phi_\lambda|}\right) ,
\end{equation}
where $v^2_{\rm th,i}\doteq 2T_{\perp\rm i}/m_{\rm i}$ is square of the (perpendicular) ion thermal speed and $\rho_{\rm th,i}\doteq v_{\rm th,i}/\Omega_{\rm i}$ is the thermal ion Larmor radius. For the change in perpendicular kinetic energy to be effective, the turbulent fluctuations must be as coherent as possible over this timescale. Denoting the typical frequency of the turbulent fluctuations at scale $\lambda$ by $\omega_\lambda$, this requirement may be written as $\omega_\lambda\tau_\lambda\sim{1}$. In this case, $\Delta K_\perp\sim q_{\rm i}\omega_\lambda \delta \Phi_\lambda\tau_\lambda \sim q_{\rm i}\delta\Phi_\lambda$.  (For a lengthier discussion of these arguments, see \S 2 and equations (12)--(16) and (24), in particular, of \citet{ChandranAPJ2010}.)

Using this information, and assuming that the stochastic gain of perpendicular kinetic energy of a single ion during the time $\tau_\lambda$ can be seen as a random walk in perpendicular-energy space, we determine the perpendicular-energy diffusion coefficient and heating rate as follows.

\subsubsection{Perpendicular diffusion coefficient and heating rate}\label{subsubsec:DQ}

We quantify the stochastic gain in an ion's perpendicular kinetic energy using the diffusion coefficient $D_{\perp\perp}^{E}\sim\Delta K_\perp^2/\tau_\lambda$. With $\Delta K_\perp \sim q_{\rm i}\delta\Phi_\lambda$ and $\tau_\lambda$ begin given by Equation \eqref{eq:taulambda}, we find
\begin{subequations}\label{eq:Dperp_Phi_general}
\begin{equation}
    D_{\perp\perp}^E(\lambda) \sim \Omega_\mathrm{i} \left(\frac{\rho_\mathrm{th,i}}{\lambda}\right)^{2} \frac{q^3_{\rm i}|\delta\Phi_\lambda|^3}{m_\mathrm{i}v_\mathrm{th,i}^2} \, . \label{eq:Dperp_Phi_general-a}
\end{equation}
Alternatively, $D_{\perp\perp}^E$ may be expressed in velocity space by using the condition $k_\perp \rho_{\rm i} = k_\perp w_\perp/\Omega_{\rm i} \sim 1$ to replace $\lambda$ with $(w_\perp/v_{\rm th,i})\rho_{\rm th,i}$. Then, denoting the resulting velocity-space potential $\delta \Phi_\lambda |_{\lambda \sim w_\perp / \Omega_{\rm i}}$ as $\delta \Phi_w$, Equation \eqref{eq:Dperp_Phi_general-a} may be reinterpreted as
\begin{equation}
    D_{\perp\perp}^E(w_\perp) \sim \Omega_\mathrm{i}\,\frac{q^3_{\rm i}|\delta\Phi_w|^3}{m_\mathrm{i}w_\perp^2} \,.\label{eq:Dperp_Phi_general-b}
\end{equation}
\end{subequations}
This equation states that particles drawn from different regions of the perpendicular distribution function experience different perpendicular energization, depending on the part of the spectrum of the fluctuations that they sample during their orbits and off of which they stochastically diffuse.

To obtain an equation for how this diffusion affects the evolution of the perpendicular-energy distribution function, $f^E$, we insert Equation \eqref{eq:Dperp_Phi_general-b} into the Fokker--Planck-like equation
\begin{equation}\label{eq:diff_equation}
    \pD{t}{f^E} = \pD{e_\perp}{} \left( D_{\perp\perp}^E\,\pD{e_\perp}{f^E}\right) ,
\end{equation}
where $e_\perp\doteq w_\perp^2/2$ is the ion's perpendicular kinetic energy per unit mass. Then, using Equation \eqref{eq:diff_equation}, we may write the total perpendicular heating as
\begin{equation}\label{eq:Qprp_Dprpprp}
    Q_\perp = -\int\mathrm{d}e_\perp\, D_{\perp\perp}^E\,\pD{e_\perp}{f^E} \, .
\end{equation}
Alternatively, one may introduce a differential heating rate in $w_\perp$ via\footnote{This definition is consistent with the diagnostics implemented in our simulations (see \S\ref{sec:simulations}). \citet{VasquezAPJ2020} argue for an alternative definition of $\partial Q_\perp/\partial w_\perp$, one which nevertheless results in the same total heating rate given by Equation \eqref{eq:Qprp_Dprpprp}. Further discussion of this alternative definition and its use in analyzing our simulation results is provided in Appendix \ref{app:sec:diffusion}.}
\begin{equation}\label{eq:dQprpdvprp_def}
    \pD{w_\perp}{Q_\perp} \doteq -D_{\perp\perp}^E(w_\perp)\, \pD{w_\perp}{f^E(w_\perp)} \,,
\end{equation}
with $D_{\perp\perp}^E(w_\perp)$ given by Equation (\ref{eq:Dperp_Phi_general-b}). 
Equation (\ref{eq:dQprpdvprp_def}) will be used in \S\ref{sec:simulations} to compute $D_{\perp\perp}^E(w_\perp)$ using the functions $f^E(w_\perp)$ and $\partial Q_\perp/\partial w_\perp$ obtained directly from our numerical simulations.

It is helpful at this stage to work through a simple estimate for how $D^E_{\perp\perp}$ and $\partial Q_\perp/\partial w_\perp$ would scale with $w_\perp$ for a particular scaling law of the fluctuating potential. Let us assume that the dominant contribution to the electric field is due to $\delta\bb{u}_{\rm i}\btimes\bb{B}/c$ induction from a fluctuating ion velocity field $\delta\bb{u}_{\rm i}$, such that $\delta\Phi_\lambda \sim  \lambda \delta u_{\perp{\rm i},\lambda}(B_0/c)$. Adopting the Kolmogorov-like scaling $\delta u_{\perp{\rm i},\lambda} \propto \lambda^{1/3}$ for these fluctuations, we find that $\delta\Phi_\lambda\propto\lambda^{4/3}$. Enacting the transformation to velocity space described above,  $\delta\Phi_w \propto (w_\perp/v_{\rm th,i})^{4/3}$. Equation \eqref{eq:Dperp_Phi_general-b} then gives $D^E_{\perp\perp} \propto (w_\perp/v_{\rm th,i})^2$, which is a scaling that matches the one of \citet{KleinChandranAPJ2016} when the induction term, ${\sim}u_{\perp{\rm i},\lambda}B_0/c$, is the dominant contribution to the electrostatic potential.\footnote{Note that \citet{KleinChandranAPJ2016} adopt $\delta u_{\perp,\lambda}\propto\lambda^{1/4}$, consistent with the dynamic-alignment argument of \citet{Boldyrev2006}. Then $\delta\Phi_\lambda\propto\lambda\delta u_{\perp,\lambda}\propto\lambda^{5/4}$ and Equation \eqref{eq:Dperp_Phi_general-b} gives $D_{\perp\perp}^E\propto w_\perp^{7/4}$, consistent with equation (17) of \citet{KleinChandranAPJ2016}.} Further assuming a Maxwellian distribution in $w_\perp$ yields a differential heating rate $\partial Q_\perp/\partial w_\perp\propto (w_{\perp}/v_\mathrm{th,i})^3 \exp(-w_\perp^2/v_\mathrm{th,i}^2)$. In this case, ion particles whose perpendicular velocities satisfy $v^2_\perp = (3/2) v^2_{\rm th,i}$ would experience the largest differential heating rate. 

\subsubsection{Exponential suppression of stochastic heating}\label{subsubsec:theory_exp_suppression}

In order to take into account the reduction of stochastic heating due to the near-conservation of the particles' magnetic moments when the fluctuations' amplitudes at the $\rho_\mathrm{th,i}$ scale are ``sufficiently small'', \citet{ChandranAPJ2010} proposed a multiplicative exponential suppression term of the type $\exp(-c_2/\xi_{\rm th})$ in Equation (\ref{eq:Dperp_Phi_general}), where $c_2$ is a (small, scale-independent) constant. This quasi-conservation condition is quantified by a so-called {\it stochasticity parameter} $\xi$, which in our theory would read as a scale-dependent parameter defined by\footnote{When the induction term provides the dominant contribution to the electrostatic fluctuations, and using the condition $\lambda \sim(w_\perp/v_\mathrm{th,i})\rho_\mathrm{th,i}$ to obtain $\delta\Phi_w$, our definition of $\xi$ reduces to (a scale-dependent version of) the definition $\xi=\delta u_{\perp}/w_\perp$ of \citet{ChandranAPJ2010}. In that work this parameter (evaluated at the ion-thermal Larmor scale) is called $\varepsilon$. However, in order to avoid confusion with the symbol typically used for the cascade rate, as well as to differentiate the generalized stochasticity parameter based on potential fluctuations from that based on ion flow-velocity fluctuations, we use $\xi$ instead. When the need arises to refer specifically to Chandran et al.'s stochasticity parameter (namely, in \S\ref{sec:interpretation}), we adopt the notation $\epsilon_{\rm i}$.\label{footnote:stochasticity}}
\begin{equation}\label{eq:xi_w_DEF}
  \xi_w\,
  \doteq\,
  \frac{q_\mathrm{i}|\delta\Phi_w|}{m_\mathrm{i}w_\perp^2}\,.
\end{equation}
The parameter $\xi_{\rm th}$, which is $\xi_w$ evaluated at the ion-thermal speed $w_\perp\sim v_\mathrm{th,i}$ (or, equivalently, at the ion-thermal gyroradius, $\lambda\sim\rho_\mathrm{th,i}$), provides an estimate of the amount of energy in the electrostatic-potential fluctuations that goes into stochastic heating, weighted by the particles' thermal energy, {\it viz.}, $\xi_\mathrm{th}\sim q_\mathrm{i}|\delta\Phi_\mathrm{th}|/m_\mathrm{i}v_\mathrm{th,i}^2$, where $\delta\Phi_\mathrm{th}$ is the velocity-space potential $\delta\Phi_w$ evaluated at $w_\perp\sim v_\mathrm{th,i}$. 
An exponential suppression factor would be justified if $\xi_\mathrm{th}\ll c_2$. 
One may then obtain a rough estimate for when the amplitude of the potential fluctuations is ``sufficiently large'' for stochastic heating to be important, that is, when the (thermal-)Larmor-scale potential satisfies $q_{\rm i}|\delta\Phi_{\rm th}|/m_{\rm i}v^2_{\rm th,i} \gtrsim c_2$. An assortment of test-particle calculations \citep{ChandranAPJ2010,XiaAPJ2013} has suggested values for $c_2$ in the range ${\approx}0.1$--$0.3$. Analyses of solar-wind data in the context of stochastic heating have adopted similar values of $c_2$~\citep{Chandran2010,Bourouaine13,MartinovicAPJ2019,MartinovicAPJS2020}.\footnote{In contrast, perpendicular ion heating measured in low-resolution hybrid-kinetic simulations of decaying Alfv\'{e}n-wave turbulence by \citet{VasquezAPJ2015} suggests that $c_2 \lesssim 0.03$, if $\xi_{\rm th}$ is calculated using the $\bb{E}\btimes\bb{B}_0$ drift evaluated on scales in the vicinity of $\rho_{\rm th,i}$.}

In our theory, we allow for an analogous, scale-dependent exponential suppression term, so that Equation \eqref{eq:Dperp_Phi_general-b} becomes (after using Equation \eqref{eq:xi_w_DEF} to replace $q_{\rm i}|\delta\Phi_w|$ with $m_{\rm i}w^2_\perp \xi_w$)
\begin{equation}\label{eq:Dperp_Phi_general_xi_EXPcorr}
   \frac{D_{\perp\perp}^E(w_\perp)}{\Omega_{\rm i}\,m_{\rm i}^{2} v_{\rm th,i}^{4}}\sim \left(\frac{w_\perp}{v_{\rm th,i}}\right)^4\xi_{w}^{3} \, \exp\left(-\frac{c_*}{\xi_{w}}\right) ,
\end{equation}
where $c_*$ is a constant to be determined. The notation $c_*$ differs from the notation $c_2$ used by \citet{ChandranAPJ2010} to emphasize that the exponential correction is being applied within the scale-dependent formulation of $D_{\perp\perp}^E(w_\perp)$, rather than within the scale-independent formulation with $w_\perp\approx v_{\rm th,i}$ (or, equivalently, $\lambda\approx\rho_{\rm th,i}$; cf.~equations (20)--(25) of \citealt{ChandranAPJ2010}).
For this reason, the value of $c_*$ does not necessarily match that of $c_2$ found in previous work.\footnote{\citet{KleinChandranAPJ2016} also allowed for a velocity-dependent exponential suppression in their formulation of $D^E_{\perp\perp}(w_\perp)$ (see their equations (8) and (17)), associating $c_*$ with $c_2 = 0.2$.} 
Within this scale-dependent formulation, a potential fluctuation is ``sufficiently large'' to heat perpendicularly an ion with velocity $w_\perp$ effectively when its amplitude satisfies $(c/B_0)|\delta\Phi_w|\gtrsim c_*\,w_\perp^2/\Omega_{\rm i}$. In terms of perpendicular scales $\lambda$, this corresponds to the range for which $(c/B_0)|\delta\Phi_\lambda|/\lambda^2\gtrsim c_*\,\Omega_{\rm i}$.
We further caution that this ``constant'' may be dependent upon $\beta$ and/or the level of intermittency in the ion-Larmor-scale fluctuations, the two possibly being related to each other as $\beta$ decreases \citep[e.g.,][]{CerriAPJL2017,GroseljAPJ2017}. 
Such intermittency could indeed partially compensate for the simultaneous decrease of $\xi_{\rm th}$ that would be associated with the enhanced separation between injection and $\rho_{\rm th,i}$ scales in the $\beta\ll1$ regime, which is precisely the regime in which stochastic heating is likely to be most relevant. This possibility seems to be supported by our simulation results (see \S\ref{subsec:heating}); future kinetic simulations with yet larger scale separations, and thus statistically smaller values of $\xi$, than those performed here are needed to investigate further the behavior of this exponential correction.

It is worth noting that, while the exponential suppression factor was originally introduced to account for the reduction in perpendicular heating when ion-Larmor-scale fluctuations are small, this factor also serves to suppress stochastic heating by larger-scale fluctuations (despite their larger relative amplitudes). Qualitatively, the lower frequencies of these fluctuations allow the ions to drift smoothly in a quasi-static potential, precluding chaotic motion and preserving approximate adiabatic invariance. Quantitatively, we may rewrite the argument of the exponential term in Equation \eqref{eq:Dperp_Phi_general_xi_EXPcorr} as $-c_* \Omega_{\rm i}\tau_w$, where $\tau_w$ is given by Equation \eqref{eq:taulambda} with $\lambda/\rho_{\rm th,i}\sim w_\perp/v_{\rm th,i}$. Then the requirement for strong suppression of stochastic heating becomes $\omega_w / \Omega_{\rm i} \ll c_* \omega_w \tau_w \lesssim c_*$, where $\omega_w$ is the frequency of gyro-scale fluctuations as seen by particles with gyro-radius $\rho_{\rm i} = w_\perp/\Omega_{\rm i} \sim \lambda$. Conversely, fluctuations whose frequencies satisfy $\omega_w/\Omega_{\rm i} \gtrsim c_*$ are the most effective at stochastically heating the ions. 

\subsection{Generalized Ohm's law and contributions to  stochastic ion heating}\label{subsec:theory_OhmLawArguments}

While the example given at the end of \S\ref{subsubsec:DQ} is illustrative, the $\delta\bb{u}_{\rm i}\btimes\bb{B}/c$ inductive electric field contributes just one piece to a more general Ohm's law. In particular, because the mechanism of stochastic ion heating occurs primarily at ion-kinetic scales (which are much smaller than the injection scales), contributions to the electric field from, e.g., the Hall effect may be important, particularly at low values of $\beta_{\rm i}$ at which the ion skin depth $d_{\rm i} \gg \rho_{\rm th,i}$. To quantify these contributions, we adopt the following generalized Ohm's law for the electric field $\bb{E}$ in which electron-inertia effects have been neglected but contributions from the Hall and thermo-electric fields are retained:
\begin{equation}\label{eq:generalized_Ohm}
\bb{E} = -\frac{\bb{u}_{\rm i}\btimes\bb{B}}{c} + \frac{\bb{J}\btimes\bb{B}}{enc} - \frac{\grad p_\mathrm{e}}{en}\,.
\end{equation}
Here we have used quasi-neutrality to replace the electron number density $n_{\rm e}$ with the ion number density $n$. Equation \eqref{eq:generalized_Ohm} is valid at scales $\lambda$ much larger than the electron-kinetic scales, {\em viz.}, $\lambda\gg d_\mathrm{e}$, $\rho_\mathrm{th,e}$, where $d_\mathrm{e}$ and $\rho_\mathrm{th,e}$ are the electron skin depth and thermal Larmor radius, respectively.\footnote{Here, we are considering scales relevant to stochastic ion heating, i.e., $k_\perp\rho_{\rm th,i}\sim1$. In our treatment, electron-inertia terms and electron finite-Larmor radius corrections can be neglected in \eqref{eq:generalized_Ohm}, if $k_\perp d_{\rm e}\ll1$ and $k_\perp\rho_{\rm th,e}\ll1$ hold at ion scales. This means that we are considering a range of $\beta_{\rm i}$ that is still larger than the (small) electron-to-ion mass ratio, i.e., $m_{\rm e}/m_{\rm i}\ll\beta_{\rm i}\lesssim1$, as well as a range of temperature ratio, $\tau_\perp$, that is smaller than the (large) inverse of such mass ratio, i.e., $0\leq \tau_\perp\ll m_{\rm i}/m_{\rm e}$.} To simplify matters further, we adopt an isothermal equation of state for the electrons, so that the electron pressure $p_{\rm e} = n T_{\rm e}$ with $T_{\rm e} = {\rm const}$. This is a good approximation for KAW fluctuations at perpendicular scales satisfying $\rho_{\rm th,e}\ll \lambda \ll \rho_{\rm th,i}$, for which the electron response is Boltzmann and therefore isothermal \citep[see, e.g., \S 7.2 of][]{SchekochihinAPJS2009}.

To obtain the potential contribution to the electric field \eqref{eq:generalized_Ohm}, we consider AW/KAW turbulence in which the fluctuations are anisotropic with respect to the magnetic-field direction, with $k_\parallel\ll k_\perp$. As in \S\ref{subsec:theory_revisited}, we therefore assume that the electric field is dominated by its potential contribution and write $|\bb{E}| \approx |\delta\bb{E}_{\perp,\lambda}| \sim \delta\Phi_\lambda / \lambda$. The other terms on the right-hand side of Equation \eqref{eq:generalized_Ohm} are then ordered as follows:
\begin{align}
    \frac{|\bb{u}_{\rm i}\btimes\bb{B}|}{c} &\sim \delta u_{\perp{\rm i},\lambda} \frac{B_0}{c} ,\\*
    \frac{|\bb{J}\btimes\bb{B}|}{enc} &\sim v_{\rm A0} \frac{d_{\rm i}}{\lambda} \left( \frac{\delta B_{\parallel,\lambda}}{B_0} + \frac{\lambda}{\ell_{\|,\lambda}} \frac{\delta B_{\perp,\lambda}}{B_0} \right) \frac{B_0}{c} , \label{eqn:Hall}\\*
    \frac{|\grad p_e|}{en} &\sim c_{\rm s} \frac{\rho_{\rm s}}{\lambda} \frac{\delta n_\lambda}{n} \frac{B_0}{c} ,
\end{align} 
where $v_\mathrm{A0}\doteq B_0/\sqrt{4\pi m_\mathrm{i}n}$ is the Alfv\'en speed, $c_{\rm s} \doteq \sqrt{T_{\rm e}/m_{\rm i}}$ is the sound speed, and $\rho_{\rm s} \doteq c_{\rm s}/\Omega_{\rm i}$ is the sound radius. 
In the Hall term (Equation \ref{eqn:Hall}), $\ell_{\|,\lambda}$ is the characteristic lengthscale along the magnetic-field direction of a fluctuation with perpendicular extent $\lambda$; the ratio $\Theta_\lambda \doteq \lambda/\ell_{\|,\lambda}$ is related to the (possibly scale-dependent) anisotropy of the turbulent cascade.

Finally, we assume that the sub-ion-scale fluctuations are composed primarily of KAWs, an assumption supported by measurements in the solar wind~\citep[e.g.,][and references therein]{ChenJPP2016}. Such a cascade satisfies approximate perpendicular pressure balance~\citep{SchekochihinAPJS2009,KunzJPP2018}: $\delta n_\lambda / n \approx -(2/\beta_\perp) \delta B_{\parallel,\lambda} / B_0$, where $\beta_\perp \doteq (1+\tau_\perp)\beta_{\perp\rm i}$. This allows one to combine the thermo-electric potential with the $\delta B_{\parallel,\lambda}$ term in the electrostatic piece of the Hall field to obtain
\begin{align}\label{eq:elec_potential_kaw}
    \frac{\delta\Phi_\lambda}{\lambda} &\sim 
    \delta u_{\perp{\rm i},\lambda} \frac{B_0}{c} 
    \nonumber\\*
    \mbox{} &+ v_{\rm A0} \frac{d_\mathrm{i}}{\lambda}\left(\frac{1}{1+\tau_\perp}\frac{\delta B_{\|,\lambda}}{B_0}
    + \Theta_\lambda\frac{\delta B_{\perp,\lambda}}{B_0}\right)\frac{B_0}{c} .
\end{align}
For a critically balanced Alfv\'{e}nic cascade with enough separation between the outer scale $L$ and $\rho_{\rm th,i}$, the spectral anisotropy $\Theta_\lambda$ becomes ${\ll}1$ as the ion-kinetic scales are approached. As a result, the contribution from the $\Theta_\lambda\delta B_{\perp,\lambda}$ term in Equation \eqref{eq:elec_potential_kaw} at a given perpendicular scale $\lambda\ll L$ may be small enough when compared to that of the field-parallel fluctuations, $\delta B_{\|,\lambda}$, to be neglected. (Note that $\delta B_\perp/\delta B_\parallel \approx\sqrt{1+2/\beta_\perp}$ for KAW-like fluctuations, e.g., see \S3.6.2 of \citealt{KunzJPP2018}.) We make this assumption in the remainder of the paper and drop the term ${\propto}\delta B_{\perp,\lambda}$ in Equation \eqref{eq:elec_potential_kaw}.\footnote{In our simulations (see \S\ref{sec:simulations}), $\Theta\approx0.05$ at $k_\perp\rho_{\rm th,i}\approx1$. This corresponds to an angle between the fluctuations' wavevector, $\bb{k}$, and the {\em local} background magnetic-field direction (i.e., using a scale-dependent definition of the background magnetic field, $\bb{B}_{\rm loc}(\bb{r},\ell)$, computed via 5-point increments; \citealt{CerriFSPAS2019}) of $\vartheta_{(\bs{k},\bs{B})}=\arctan(\Theta^{-1})\approx87^\circ$. We note that $\vartheta_{(\bs{k},\bs{B})} \approx 80^\circ$--$90^\circ$ for fluctuations measured in the near-Earth solar wind with spacecraft-frame frequencies $f_{\rm spacecraft} \sim 1~{\rm Hz}$ \citep{SahraouiPRL2010}.}

Converting Equation \eqref{eq:elec_potential_kaw} without the $\Theta_\lambda$ term into the velocity space potential $\delta\Phi_w$ and inserting it in Equation \eqref{eq:Dperp_Phi_general-b} (i.e., neglecting the multiplicative exponential suppression factor in Equation \eqref{eq:Dperp_Phi_general_xi_EXPcorr} for the moment), one obtains an analytic formula for the perpendicular-energy diffusion coefficient,
\begin{equation}\label{eq:Dperp_Phi_asymptotic}
    \frac{D_{\perp\perp}^E}{\Omega_\mathrm{i}m_\mathrm{i}^2v_\mathrm{th,i}^4}
    \sim 
    \left(\frac{w_\perp}{v_\mathrm{th,i}}\right)
    \left(
    \frac{\delta u_{\perp{\rm i},w}}{v_\mathrm{th,i}}
    +
    \frac{1}{\beta_\perp}\,\frac{v_\mathrm{th,i}}{w_\perp}\,\frac{\delta B_{\|,w}}{B_0}
    \right)^3\,.
\end{equation}
Equation \eqref{eq:Dperp_Phi_asymptotic} implies that, depending on the spectral slopes of the fluctuation spectra at the ion gyro-radii, ions with different perpendicular energies will diffuse differently in velocity space. This dependence is computed in \S\ref{subsec:theory_AW-KAW_scaling}, where we assign various spectral scaling laws to $\delta u_{\perp{\rm i},\lambda}$ and $\delta B_{\parallel,\lambda}$ that correspond to different regimes of AW/KAW turbulence. These are then substituted into Equation \eqref{eq:elec_potential_kaw} with $\lambda \sim (w_\perp/v_{\rm th,i}) \rho_{\rm th,i}$, thereby yielding the velocity-scale dependence of $\delta \Phi_w$ and, through Equations \eqref{eq:Dperp_Phi_general-b} and \eqref{eq:Qprp_Dprpprp}, $D^E_{\perp\perp}$ and $Q_\perp$. In preparation for this exercise, we first advance arguments for which of the terms in Equation \eqref{eq:elec_potential_kaw} provides the dominant contribution to the potential as seen by a particle with Larmor radius $\rho_\mathrm{i}$ (when compared to the thermal gyro-radius, $\rho_\mathrm{th,i}$, and to the ion skin depth, $d_\mathrm{i}$ -- and thus depending upon $\beta_\perp$ as well).

Stochastic heating of an ion with perpendicular random velocity $w_\perp$ involves fluctuations that occur on scales comparable to that ion's gyro-radius, $\lambda\sim\rho_\mathrm{i} = w_\perp/\Omega_\mathrm{i}$. 
This scale must be compared with the ion-kinetic scales of the background plasma, namely $\rho_\mathrm{th,i}$ and $d_\mathrm{i}$, which determine the nature of the turbulent fluctuations at scale $\lambda$ and thus the corresponding ordering of the different terms in Equation~\eqref{eq:elec_potential_kaw}. 
These background spatial scales also have a corresponding scale in perpendicular velocity, namely the ion-thermal and Alfv\'en speeds, $v_\mathrm{th,i}=\Omega_\mathrm{i}\rho_\mathrm{th,i}$ and $v_\mathrm{A}=\Omega_\mathrm{i}d_\mathrm{i}=v_{\rm th,i}/\sqrt{\beta_{\perp\rm i}}$, respectively.
Just as the spatial scales determine the type of fluctuations that are responsible of the stochastic heating, these background velocity scales -- and how they compare with the ion's velocity $w_\perp$ -- determine the corresponding ordering of the different terms in Equation~\eqref{eq:Dperp_Phi_asymptotic}. 
Moreover, as discussed in \S\ref{subsec:theory_revisited}, for a quasi-Maxwellian distribution we expect that the largest contribution to the total stochastic heating is provided by ions with $w_\perp\sim v_\mathrm{th,i}$. 
The contribution from those ions whose perpendicular velocity exceeds a few times the ion-thermal speed, $w_\perp\gg v_\mathrm{th,i}$, is  exponentially suppressed. Similarly, the contribution from low-$w_\perp$ ions (i.e.~those with $w_\perp\ll v_{\rm th,i}$) to the overall heating would be progressively less important due to the strong dependence of $D_{\perp\perp}^E \propto\delta\Phi_w^3$ on the fluctuations' amplitudes ({\em viz.}, the lower the $w_\perp$, the smaller the spatial scale $\lambda\sim\rho_{\rm i}\propto w_\perp$ at which the potential is sampled). Therefore, based on these arguments and what we know about the cascade of Alfv\'enic fluctuations, we may anticipate the following features of stochastic heating in the different $\beta_\perp$ regimes. 

We first consider Equation~\eqref{eq:Dperp_Phi_asymptotic} at $w_\perp\approx v_{\rm th,i}$. When $\beta_{\perp\rm i}\gtrsim1$, we have $\rho_\mathrm{th,i}\gtrsim d_\mathrm{i}$, and so the ion thermal gyro-radius is encountered sooner by the cascading fluctuations than is the ion skin depth. At such scale, the incompressive AW-like $\delta u_{\perp{\rm i}}$ fluctuations are still dominant over their compressive KAW-like $\delta B_\|$ counterparts~\citep[e.g.,][]{CerriJPP2017,CerriAPJL2017} (which are also suppressed by an additional factor $\beta^{-1}_\perp$ in Equation \eqref{eq:Dperp_Phi_asymptotic} when $\beta_\perp > 1$). As a result, for $\beta_{\perp\rm i}\gtrsim1$, we expect that the main contribution to the overall stochastic heating of ions is provided by the potential associated with the inductive term in Equation~\eqref{eq:generalized_Ohm}.

On the other hand, if $\beta_{\perp\rm i}\ll1$, then the ion thermal Larmor radius is much smaller than the ion skin depth, $\rho_\mathrm{th,i}\ll d_\mathrm{i}$, and turbulent fluctuations encounter $d_\mathrm{i}$ as the first ion-kinetic scale in their cascade. Because the ions decouple from the dynamics of the magnetic field at sub-$d_{\rm i}$ scales, the spectrum of ion-flow-velocity fluctuations becomes much steeper than its magnetic counterpart (an effect captured by the $\bb{J}\btimes\bb{B}/en$ Hall term in Equation (\ref{eq:generalized_Ohm})). 
Accordingly, $\delta u_{\perp\rm i}$ fluctuations are negligibly small at $\lambda \ll d_\mathrm{i}$ relative to magnetic-field fluctuations, a feature that has been seen in both {\it in situ} measurements of solar-wind turbulence \cite[e.g.,][]{SafrankovaApJ2016,ChenBoldyrevAPJ2017} and in kinetic numerical simulations of Alfv\'{e}nic turbulence~\citep[e.g.,][]{CerriJPP2017,FranciAPJ2018,ArzamasskiyAPJ2019}. 
Moreover, at $\beta_\perp<1$, the compressive KAW-like $\delta B_\|$ contribution to Equation  \eqref{eq:Dperp_Phi_asymptotic} is now further enhanced by the factor $\beta^{-1}_\perp$.
As a result, in the low-$\beta$ regime, we anticipate the main contribution to the overall stochastic heating of ions to be provided by the potential associated with the non-ideal terms in Equation \eqref{eq:generalized_Ohm}.

\subsection{Explicit scalings for stochastic ion heating from a critically balanced, Alfv\'enic cascade}\label{subsec:theory_AW-KAW_scaling}

In this section we utilize well-known spectral scaling relations for $\delta u_{\perp{\rm i},\lambda}$ and $\delta B_{\parallel,\lambda}$ in AW and KAW turbulence to evaluate Equation \eqref{eq:Dperp_Phi_asymptotic} and the associated perpendicular heating rate, Equation \eqref{eq:Qprp_Dprpprp}. To keep our expressions compact, we neglect for the time being the exponential suppression factor. A brief comment on how this factor modifies the results is then provided in \S\ref{subsubsec:exp_attenuation}; the full calculation with the factor included is reported in Appendix~\ref{app:sec:exact_calculation_Qperp}. Strictly speaking, the contents of this section (\S\ref{subsec:theory_AW-KAW_scaling}) are not fully self-consistent, in that the transfer of turbulent energy to the thermal energy of the particles via stochastic heating is not accounted for in the adopted spectral scalings (which are power-law in form). However, it does allow us to gain some intuition for how $D_{\perp\perp}$ might scale with $w_\perp$ and how the perpendicular heating rate per unit mass $Q_\perp$ depends on the plasma parameters. In doing so, we are most closely following \citet{KleinChandranAPJ2016}, who noted that their approach neglects the back reaction of the heating process on the turbulent power spectrum. A self-consistent determination of $D_{\perp\perp}$ and $Q_\perp$ follows in \S\ref{sec:simulations}, where we obtain spectral scalings for $\delta\Phi_\lambda$, $\delta u_{\perp{\rm i},\lambda}$, and $\delta B_{\parallel,\lambda}$ from self-consistent numerical simulations and use them in Equations \eqref{eq:dQprpdvprp_def}, \eqref{eq:Dperp_Phi_general_xi_EXPcorr}, and \eqref{eq:Dperp_Phi_asymptotic} to determine $D_{\perp\perp}$ and $\partial Q_\perp/\partial w_\perp$.

Consider an inertial-range cascade of large-scale (MHD) Alfv\'{e}nic fluctuations characterized by a constant energy cascade rate per unit mass $\varepsilon_{\rm AW}$ and $\delta u_{\perp{\rm i},\lambda} \sim (\varepsilon_{\rm AW} \lambda)^{1/3}$. This cascade is taken to exhibit a scale-dependent spectral anisotropy governed by critical balance \citep{GoldreichSridharAPJ1995,HorburyPRL2008}, such that the characteristic field-parallel lengthscale of a fluctuation of perpendicular size $\lambda$ satisfies $\ell_{\parallel,\lambda} \sim L^{1/3} \lambda^{2/3}$, where $L \doteq v^3_{\rm A} / \varepsilon_{\rm AW}$ is the outer scale. As the ion kinetic scales are approached, the AWs mutate into KAWs, with a fraction $\varepsilon_{\rm KAW}/\varepsilon_{\rm AW}$ of the inertial range cascade energy penetrating down into the dispersive range.\footnote{In gyrokinetic turbulence, the AW energy that does not make its way into the KAW cascade channel while going through the ion kinetic scales is transferred into ion thermal energy through Landau damping and/or a perpendicular phase-space cascade of ion-entropy fluctuations \citep{SchekochihinAPJS2009}. Here, we allow for a portion of the cascading energy to go also into perpendicular stochastic heating of the ions. According to the discussion that follows Equation \eqref{eq:Dperp_Phi_asymptotic} in \S\ref{subsec:theory_OhmLawArguments}, when $\beta_\perp\gtrsim{1}$ this heating mechanism drains a portion of the energy carried by the AW cascade ($\varepsilon_{\rm AW}$), while it is a portion of the KAW cascade ($\varepsilon_{\rm KAW}$) that is going into such ion-energy channel at $\beta\ll1$.} For the sub-ion-scale KAW cascade, we do not adhere to any particular prescription for the associated wavevector anisotropy, using instead a generalized version of equation (4.47) of \citet{KunzJPP2018},
\begin{equation}\label{eq:KAWanisotropy_alpha}
    \ell_{\|,\lambda}\sim
    \left(\frac{\varepsilon_\mathrm{AW}}{\varepsilon_\mathrm{KAW}}\right)^{1/3}
    \frac{(1+\tau_\perp)^{1/6}}{(2+\beta_\perp)^{1/6}}\,L^{1/3}\,\rho_\mathrm{th,i}^{2/3}\left(\frac{\lambda}{\rho_\mathrm{th,i}}\right)^{\alpha/3}\,,
\end{equation}
in which the anisotropy is parametrized by the exponent $\alpha$~\citep{CerriAPJL2018}. Different values of $\alpha$ may result by assuming different non-linear energy transfer timescales that govern the critically balanced cascade. For example, $\alpha = 1$ corresponds to a conservative KAW cascade with spectral slope $-7/3$, as predicted by the gyrokinetic theory  \citep[e.g.,][]{SchekochihinAPJS2009}. Accounting for a scale-dependent volume-filling factor of the KAW fluctuations instead yields $\alpha=2$, with an associated KAW spectrum having a slope of $-8/3$ \citep{BoldyrevPerezAPJL2012}. Finally, $\alpha=3$ corresponds to a scale-independent anisotropy, a feature sometimes seen in hybrid-kinetic simulations of AW/KAW turbulence \citep[e.g.,][]{FranciAPJ2018,ArzamasskiyAPJ2019} and predicted by theories of reconnection-mediated Alfv\'enic turbulence \citep{LoureiroBoldyrevAPJ2017,MalletMNRAS2017}.

\subsubsection{Stochastic heating in $\beta\gtrsim1$ AW turbulence}\label{subsubsec:theory_AWscaling_explicit}
 
When $\beta_\mathrm{\perp i}\gtrsim1$, the nonlinear fluctuations approaching the ion Larmor scale are composed primarily of AWs. Therefore, the main contribution to the electrostatic potential in (\ref{eq:elec_potential_kaw}) is from the $\delta u_{\perp\rm i}$ fluctuations, and the diffusion coefficient can be approximated by
\begin{equation}\label{eq:Dperp_Phi_AWlimit}
    \frac{D_{\perp\perp}^E}{\Omega_\mathrm{i}m_\mathrm{i}^2v_\mathrm{th,i}^4}
    \approx
    \left(\frac{w_\perp}{v_\mathrm{th,i}}\right)
    \left(\frac{\delta u_{\perp{\rm i},w}}{v_\mathrm{th,i}}\right)^3 ,
\end{equation}
with the Alfv\'enic fluctuations satisfying
\begin{equation}\label{eq:dUperp_scaling_AW}
    \frac{\delta u_{\perp{\rm i},\lambda}}{v_\mathrm{th,i}}
    \sim \left(\frac{\varepsilon_\mathrm{AW}}{\Omega_{\mathrm{i}}v_\mathrm{A0}^2}\right)^{1/3}\hspace{-0.1pt}
    \beta_{\perp\rm i}^{-1/3}
    \left(\frac{\lambda}{\rho_\mathrm{th,i}}\right)^{1/3} .
\end{equation}
Substituting this expression into \eqref{eq:Dperp_Phi_AWlimit} with $\lambda/\rho_{\rm th,i} \sim w_\perp / v_{\rm th,i}$ yields
\begin{equation}\label{eq:Dperp_Phi_scaling_AW}
    D_{\perp\perp}^\mathrm{(AW)} \sim \varepsilon_\mathrm{AW}\,m_\mathrm{i}^2v_\mathrm{th,i}^2\left(\frac{w_\perp}{v_\mathrm{th,i}}\right)^2.
\end{equation}
By using Equation (\ref{eq:Qprp_Dprpprp}) and adopting for simplicity a Maxwellian distribution function in $w_\perp$, $f^E(w_\perp)=\exp(-w_\perp^2/v_\mathrm{th}^2)/(m_\mathrm{i}v_\mathrm{th}^2)$, we find that the perpendicular heating rate per unit mass is given by
\begin{equation}\label{eq:Qperp_Phi_scaling_AW}
    \frac{Q_\perp^\mathrm{(AW)}}{\varepsilon_{\rm AW}} = \Lambda_\mathrm{AW} ,
\end{equation}
where $\Lambda_\mathrm{AW}$ is a constant independent of $\beta_{\perp\rm i}$ and $\tau_\perp$ that takes into account the various coefficients neglected in our scaling arguments. Therefore, at any $\beta_\perp\gtrsim1$ the stochastic-heating rate (associated to AW-like fluctuations only) obtains an approximately constant fraction of the energy cascade rate. This result is consistent with the one in \citet{ChandranAPJ2010} for the case in which the dominant contribution to the electric-field fluctuations is due to the $\delta\bb{u}_{\rm i}\btimes\bb{B}_0/c$ induction (and the exponential suppression factor is neglected; cf.~their equation 31).


\subsubsection{Stochastic heating in low-$\beta$ KAW turbulence}\label{subsubsec:theory_KAWscaling_explicit}

When $\beta_\perp\ll 1$, the ion Larmor radius is smaller than the ion skin depth, $\rho_\mathrm{i} \sim \rho_{\rm th,i}\ll d_\mathrm{i}$. As a result, the fluctuating potential \eqref{eq:elec_potential_kaw} evaluated at ion-Larmor scales is dominated by the contribution from the $\delta B_{\parallel,\lambda}$ fluctuations, and the diffusion coefficient can be approximated by
\begin{equation}\label{eq:Dperp_Phi_KAWlimit}
    \frac{D_{\perp\perp}^E}{\Omega_\mathrm{i}m_\mathrm{i}^2v_\mathrm{th,i}^4}\,
    \approx\, 
    \beta_\perp^{-3}\,
    \left(\frac{w_\perp}{v_\mathrm{th,i}}\right)^{-2}
    \left(\frac{\delta B_{\|,w}}{B_0}\right)^3\,, 
\end{equation}
with compressive KAW-like fluctuations satisfying
\begin{equation}\label{eq:dBpara_scaling_KAW}
    \frac{\delta B_{\|,\lambda}}{B_0}
    \sim 
    \left(\frac{\varepsilon_\mathrm{KAW}}{\Omega_{\mathrm{i}}v_\mathrm{A0}^2}\right)^{1/3} \frac{\beta^{1/3}_{\perp\rm i}}{(1+2/\beta_\perp)^{1/3}}
    \left(\frac{\lambda}{\rho_\mathrm{th,i}}\right)^{(3+\alpha)/6} .
\end{equation}
Substituting this expression into \eqref{eq:Dperp_Phi_KAWlimit} with $\lambda/\rho_{\rm th,i} \sim w_\perp/v_{\rm th,i}$ yields
\begin{equation}\label{eq:Dperp_Phi_scaling_KAW}
    D^{\rm (KAW)}_{\perp\perp} \sim \varepsilon_{\rm KAW} \, m_{\rm i}^2 v^2_{\rm th,i} \frac{(1+\tau_\perp)^{-2}}{(2+\beta_\perp)} \left(\frac{w_\perp}{v_{\rm th,i}}\right)^{(\alpha-1)/2}.
\end{equation}
For $\alpha=1$, $D^{\rm (KAW)}_{\perp\perp}$ is independent of $w_\perp$; for $\alpha=2$, $D_{\perp\perp}^\mathrm{(KAW)}\propto w_\perp^{1/2}$; and for $\alpha=3$, $D_{\perp\perp}^\mathrm{(KAW)}\propto w_\perp$. Again adopting a Maxwellian distribution function in $w_\perp$, we may estimate the perpendicular heating rate per unit mass in low-$\beta$ KAW turbulence as
\begin{equation}\label{eq:Qperp_Phi_scaling_KAW}
    \frac{Q^{\rm (KAW)}_\perp}{\varepsilon_{\rm KAW}} = \Lambda_{\rm KAW} \, (1+\tau_\perp)^{-2} (2+\beta_\perp)^{-1} ,
\end{equation}
where $\Lambda_{\rm KAW}$ is a constant independent of $\beta_{\perp\rm i}$ and $\tau_\perp$.

If we further make the assumption that the transition from the AW cascade to the KAW cascade occurs at and is continuous across $k_\perp\rho_{\rm th,i} \sim 1$, then we may estimate $\varepsilon_{\rm KAW}/\varepsilon_{\rm AW} \sim (\tau_{\rm AW}/\tau_{\rm KAW})_{k_\perp \rho_{\rm th,i}\sim 1} \sim  (2+\beta_\perp)^{-1/2} (1+\tau_\perp)^{1/2}$, in which case
\begin{equation}
     \frac{Q^{\rm (KAW)}_\perp}{\varepsilon_{\rm AW}} \propto  (1+\tau_\perp)^{-3/2} \, (2+\beta_\perp)^{-3/2}.
\end{equation}
If instead the transition were to occur at $k_\perp d_{\rm i} \sim 1$ \citep[e.g.,][]{Chen2014}, then
\begin{equation}
    \frac{Q^{\rm (KAW)}_\perp}{\varepsilon_{\rm AW}} \propto \beta^{1/2}_\perp\, (1+\tau_\perp)^{-2} \, (2+\beta_\perp)^{-3/2}.
\end{equation}

\subsubsection{Exponential attenuation}\label{subsubsec:exp_attenuation}

As forewarned at the start of \S\ref{subsec:theory_AW-KAW_scaling}, we have been omitting the exponential suppression factor introduced in Equation \eqref{eq:Dperp_Phi_general_xi_EXPcorr} to keep the limiting expressions for $D^E_{\perp\perp}$ and $Q_\perp$ in different $\beta$ regimes compact. When this correction is included, the diffusion coefficient $D_{\perp\perp}^E(w_\perp)$ acquires a peak at a certain velocity, $w_\perp^{\rm(peak)}$, corresponding to the ``most affected'' (or ``quasi-resonant'') ion population. For example, an exponentially corrected diffusion coefficient of the form $D_{\perp\perp}^E(w_\perp)\propto (w_\perp/v_{\rm th,i})^a\exp[-c\,(w_\perp/v_{\rm th,i})^b]$, with constants $a,b,c\geq0$, displays a peak at perpendicular velocity $=(a/bc)^{1/b}\, v_{\rm th,i}$ (except in the case of standard KAW anisotropy, for which $\alpha=1$, $a=0$, and $D_{\perp\perp}^E$ is just an exponentially decreasing function of $w_\perp$). If the exponential suppression were important, then the differential perpendicular-heating rate, $\partial Q_\perp/\partial w_\perp$, would also peak, at $w_\perp^{\rm(peak)}= ((a+1)/bc)^{1/b}\, v_{\rm th,i}$. This would result in stochastic heating occurring most strongly on length scales $\lambda^{\rm(peak)}\approx(w_\perp^{\rm(peak)}/v_{\rm th,i})\rho_{\rm th,i}$.
On the other hand, if the fluctuations are in a regime in which the exponential correction is not important, then, to the lowest order, we recover the cases discussed in \S\ref{subsubsec:theory_AWscaling_explicit} and \S\ref{subsubsec:theory_KAWscaling_explicit}, {\em viz.}, a power-law diffusion coefficient of the form $D_{\perp\perp}^E(w_\perp)\propto (w_\perp/v_{\rm th,i})^a$, and a differential heating $\partial Q_\perp/\partial w_\perp$ peaking at $w_\perp^{\rm(peak)}=\sqrt{(a+1)/2\,}\,v_{\rm th,i}$ because of the $\partial f^E/\partial w_\perp$ factor. We refer the reader to Appendix~\ref{app:sec:exact_calculation_Qperp} for details.

\begin{figure*}[!ht]
    \centering
    \includegraphics[width=\textwidth]{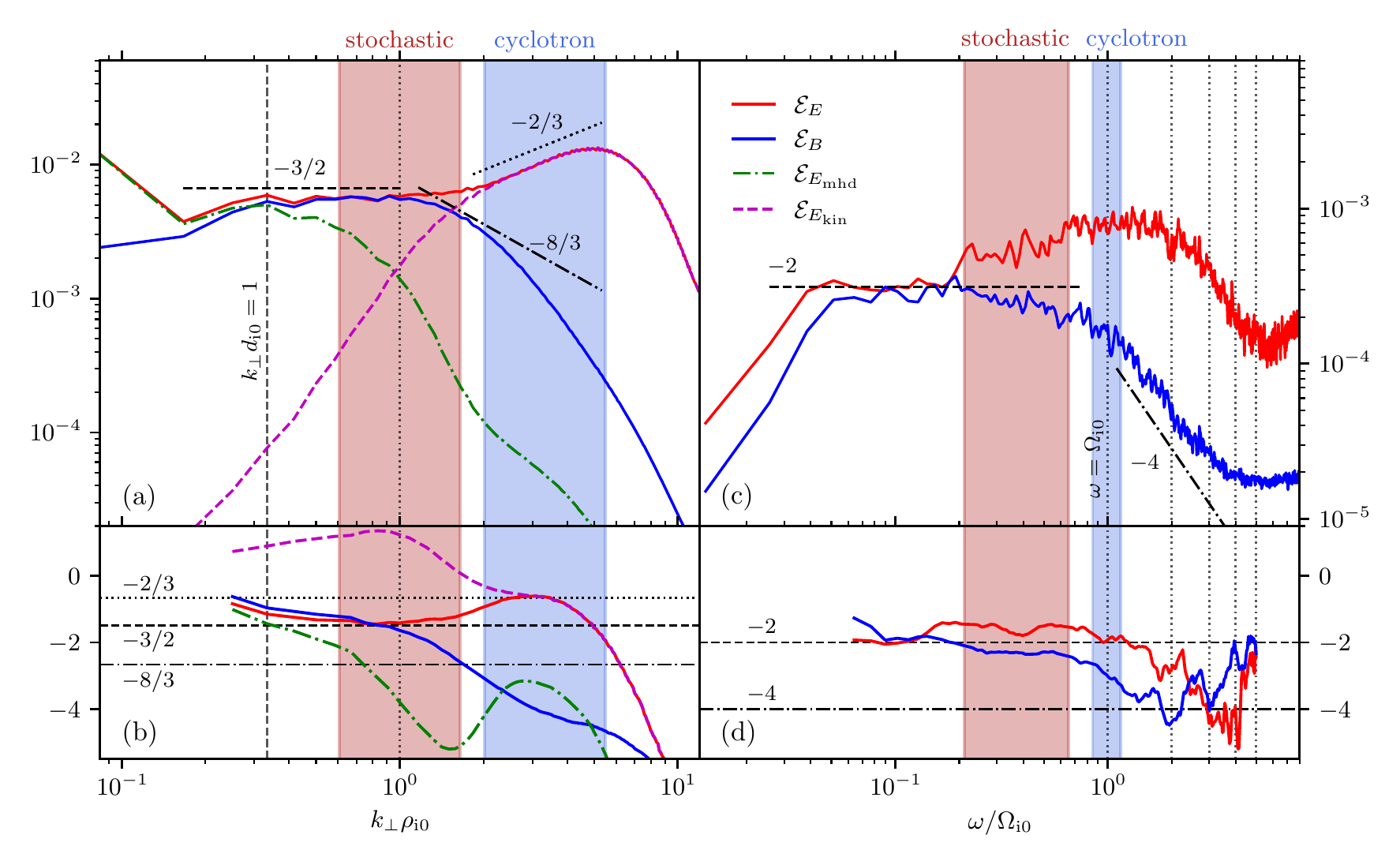}
    \caption{Compensated energy spectra (top panels) and local spectral slopes (bottom panels) for $\beta_{\mathrm{i}0} = 1/9$ simulation. (a) Wavenumber spectra (compensated by $(k_\perp\rho_{\rm i0})^{3/2}$) of magnetic field $\bb{B}$ (blue), electric field $\bb{E}$ (red), the ``MHD'' component of the electric field $\bb{E}_\mathrm{mhd}=-\bb{u}_{\rm i}\btimes\bb{B}/c$ (green dashed), and the ``kinetic'' component of the electric field $\bb{E}_\mathrm{kin}=(\bb{J}\btimes\bb{B}/c-T_\mathrm{e}\grad n)/en$ (purple dashed). (b) Local spectral slopes versus $k_\perp\rho_{\mathrm{i}0}$. (c) Frequency spectra (compensated by $(\omega/\Omega_{\rm i0})^2$) of $\bb{B}$ and $\bb{E}$ (blue and red, respectively). (d) Local spectral slopes versus $\omega/\Omega_{\mathrm{i}0}$. The light-red (light-blue) shaded region highlights the wavenumber/frequency ranges where stochastic (cyclotron) heating is considered to be important.}
    \label{fig:spectra}
\end{figure*}

\section{Numerical verification}\label{sec:simulations}

We test the theory presented in \S\ref{sec:theory} using hybrid-kinetic simulations with the particle-in-cell code \pegpp (\citealt{KunzJCP2014}; Arzamasskiy et al., {\em in prep.}). Our hybrid model consists of fully kinetic ions coupled to a massless, charge-neutralizing, isothermal electron fluid via the generalized Ohm's law \eqref{eq:generalized_Ohm}~\citep[see][for the model equations]{ArzamasskiyAPJ2019}. While hybrid-kinetics excludes electron kinetic effects such as electron Landau damping~\citep[e.g.,][]{TenBargeAPJL2013,told16b,GroseljAPJ2017}, it retains certain ion-energization mechanisms (such as stochastic heating and ion-cyclotron resonances) that are not included in other models often used to study turbulent dissipation in collisionless plasmas~\citep[e.g., gyrokinetics;][]{HowesJGR2008,ToldPRL2015,KawazuraPNAS2019}. 
We refer the interested reader to \citet{told16b} and \citet{CamporealeBurgessJPP2017} for a comparison of linear modes in hybrid-kinetics, gyrokinetics, and full kinetics.
Similarly, a comparative study of fluctuations' properties in 3D hybrid- and full-kinetic turbulence at sub-ion scales can be found in \citet{CerriFSPAS2019}.

\subsection{Simulation setup}\label{subsec:sim_setup}

We consider an initially uniform plasma with ion density $n_0$, threaded by a uniform background magnetic field $\bb{B}_0 = B_0 \bb{e}_z$ and placed within a three-dimensional, periodic computational domain of size $L_\perp^2 \times L_z$ with $L_x = L_y \doteq L_\perp$. Turbulence is driven continuously in this plasma via a random, incompressible external force $\bb{F}_\mathrm{ext}$, which excites ion momentum fluctuations in the $x$-$y$ plane perpendicular to $\bb{B}_0$. The forcing is time de-correlated over the interval $\tau_{\rm corr}$ using an Ornstein--Uhlenbeck process~\citep[see][\S 2]{ArzamasskiyAPJ2019}. Only the largest-scale modes with $k_\parallel^{F} = 2\pi/L_\parallel$ and $k_\perp^{F} = [1,2]\times2\pi/L_\perp$ are driven. Critical balance of the largest scale fluctuations is assured by choosing a forcing amplitude such that the root-mean-square (rms) mean velocity fluctuation, $u_{\rm rms}$, satisfies $u_{\rm rms} / v_{\rm A0} \approx L_\perp / L_\|$ in the quasi-steady turbulent state. Accordingly, $\tau_{\rm corr} = L_\perp/2\pi u_{\rm rms} \approx L_\|/2\pi v_{\rm A0}$ is proportional to the Alfv\'en crossing time $\tau_{\rm A}=L_\|/v_{\rm A0}$. 
At the smallest scales, dissipation of turbulent energy is achieved by means of a fourth-order hyper-resistivity on the magnetic field and low-pass filters on the first two moments of $f_{\rm i}$ ({\em viz.}, $n_{\rm i}$ and $n_{\rm i}\bb{u}_{\rm i}$).

In this paper, we combine results from two simulations of low-$\beta$ turbulence: a simulation with $\beta_{\mathrm{i}0}=0.3$ presented by \citet{ArzamasskiyAPJ2019}, and a new simulation with $\beta_{\mathrm{i}0} = 1/9$. 
This new simulation employs an elongated box with $L_\parallel = 6L_\perp = 48\pi d_{\mathrm{i}0} = 144 \pi \rho_{\mathrm{i}0}$, discretized into $N_x=N_y=288$ and $N_z=1728$ cells, achieving an isotropic resolution $\Delta x \simeq 0.087d_{\mathrm{i}0}$ (${\simeq}0.26\rho_{\mathrm{i}0}$). The simulated wavenumber space is then $0.25\leq k_\perp d_{\mathrm{ i}0}\leq 36$ and $0.04\lesssim k_\| d_{\mathrm{i}0}\leq 36$ (corresponding to $0.083\lesssim k_\perp\rho_{\mathrm{i}0}\leq 12$ and $0.014\lesssim k_\|\rho_{\mathrm{i}0}\leq 12$). In each cell, the initial ion distribution function is represented with $512$ particles (giving ${\approx}\,73$ billion particles in total).\footnote{The $\beta_{\rm i0}=0.3$ simulation of \citet{ArzamasskiyAPJ2019} had utilized a $\delta f$-method to reduce the impact of particle noise on the fluctuations. This new $\beta_{\rm i0}=1/9$ simulation adopts a full-$f$ scheme in order to better handle potentially strong local density variations that arise in this low-beta regime.} We run this simulation for ${\approx}7.6\,\tau_{\rm A}$, with the quasi-steady state developing around ${\approx}4.3\,\tau_\mathrm{A}$. Our results are time-averaged over the remaining ${\approx}3.3\,\tau_\mathrm{A}$ (corresponding to ${\approx}500\,\Omega_{\rm i0}^{-1}$). For the $\beta_{\rm i0}=0.3$ run, we define the quasi-steady state as starting from $t/\tau_{\rm A}\approx 4.4$ and continuing to the end of the simulation at $t/\tau_{\rm A}\approx 19.9$ (corresponding to ${\approx}4110\,\Omega^{-1}_{\rm i0}$).

\subsection{Fluctuation spectra for $\beta_{\rm i0}=1/9$}\label{subsec:spectra}

Figure~\ref{fig:spectra} presents energy spectra and scale-dependent spectral indices (``local slopes'') for the $\beta_{\mathrm{i}0} = 1/9$ run versus (a,b) the wavenumber $k_\perp$ perpendicular to $\bb{B}_0$ and (c,d) the frequency $\omega$ measured in the plasma frame. These fluctuations exhibit significantly different spectra than in the corresponding $\beta\sim1$ case~\citep[e.g., see][and references therein]{CerriFSPAS2019}. 
First, the MHD-range spectra of electric and magnetic fluctuations both show a slope shallower than the usual anisotropic-MHD $-5/3$ scaling~\citep[e.g.,][]{GoldreichSridharAPJ1995} and closer to $-3/2$. (This may be due to the limited scale separation between the driving scales and the ion skin depth.) Second, while the spectral slope of the electric-field energy in the kinetic range is extremely close to $-2/3$, the corresponding magnetic-field spectrum steepens continuously beyond the $-8/3$ predicted to accompany the $-2/3$ electric spectrum.

We interpret this sub-ion-Larmor steepening as a signature of energy dissipation due to {\it ion-heating mechanisms}. This interpretation is supported by the frequency spectra in Figure~\ref{fig:spectra}(c), which exhibit slopes close to the $-2$ corresponding to a conservative energy cascade at frequencies $\omega/\Omega_{\rm i0}\lesssim 0.2$, but which steepen progressively through the sub-ion-Larmor range. As we will show in \S\ref{subsec:heating}, there are two ion-heating mechanisms operating simultaneously in this range, namely {\it stochastic} and {\it cyclotron} heating. The corresponding approximate wavenumber ranges in which one of these mechanisms is measured to be dominant over the other one are indicated in Figure~\ref{fig:spectra}(a,b) as light-red (light-blue) shaded regions for stochastic (cyclotron) heating. These ranges have been determined via direct measurement of the ions' perpendicular heating versus $k_\perp$, which shows a first peak around $k_\perp\rho_{\rm i0}\sim1$ that we associate with stochastic heating and a second peak around $k_\perp\rho_{\rm i0}\sim3$ that we associate with cyclotron heating (see Figure~\ref{fig:Qprp_vs_kprp} and accompanying discussion in \S\ref{subsec:heating}). Although there would likely be an overlap between the actual ranges over which these mechanisms operate at sub-ion scales, for the sake of clarity the extent of these regions in Figure~\ref{fig:spectra}(a,b) is taken to be between $k_0/\sqrt{{\rm e}}$ and $k_0\sqrt{{\rm e}}$ ($k_0$ being the peak-wavenumber of each mechanism), a range previously used to estimate the total amount of stochastic ion heating~\citep[see, e.g.,][]{XiaAPJ2013,MartinovicAPJS2020}. The highlighted wavenumber ranges also have corresponding frequency ranges, highlighted in panels (c) and (d). These frequency ranges are obtained using an approximate AW/KAW dispersion relation for the stochastic-heating range\footnote{Namely, $\omega^2=k_\|^2v_\mathrm{A}^2\big[1+(1+\tau_\perp)k_\perp^2\rho_\mathrm{i}^2/(2+\beta_\perp)\big]$~\citep[this formula smoothly interpolates between the AW and the KAW limits; cf. eqs.(4)--(5) in][]{HowesJGR2008}. Different approximations for the KAW limit~\citep[see, e.g.,][]{LysakLotkoJGR1996} provide similar qualitative results, {\em viz.}, that $\omega\simeq\Omega_\mathrm{i}$ at $k_\perp\rho_\mathrm{i}\approx3$.} and, for cyclotron heating associated to the $n=1$ resonance, considering a resonance broadening of roughly $\Delta\omega/\omega_0\sim1/k_0\rho_{\rm i0}$ (light-blue region in panels (c) and (d)). We mention that there are also higher-$n$ resonances (shown as vertical dotted lines), likely contributing to the overall cyclotron heating.\footnote{The $n>1$ resonances are not formally associated to KAW-like fluctuations, but rather to other type of fluctuations being relevant at low $\beta$~\cite[see, e.g.,][]{CerriAPJL2016,CerriAPJL2017,GroseljAPJ2017}.}
A detailed analysis of the fluctuations' spectral features, structure functions, and turbulence-related dynamics (e.g., magnetic reconnection) will be reported on elsewhere.

Before providing diagnostic evidence supporting this claim -- that the ion- and sub-ion-Larmor-scale spectral steepening we observe is attributable to particle energization via stochastic and cyclotron heating -- we note that such an association between changes in spectral slopes and energy dissipation is a relatively old idea in the solar-wind context \citep{Coleman1968}, one that continues to be employed today \citep[e.g.,][]{WoodhamAPJ2018}. Indeed, the steepness of the magnetic spectrum has been shown to correlate with both the energy cascade rate and power level in the inertial range \citep{SmithApJL2006,BrunoAPJL2014} and the thermal proton temperature \citep{LeamonJGR1998}. A more recent example may be found in figure 5 of \citet{ChenNatCo2019}, which shows a gradual steepening of the magnetic-field power spectrum in the Earth's magnetosheath throughout the sub-ion-Larmor range. While this kind of steepening has been attributed in some theoretical models to electron Landau damping \citep{SahraouiPRL2009,HowesPOP2011,TenBargeAPJ2013,PassotAPJL2015}, the resemblance between our Figure \ref{fig:spectra}(b) and Figure 5(b) of \citet{ChenNatCo2019} is notable given that our simulations do not include electron kinetics.

\subsection{Ion heating in low-$\beta$ turbulence}\label{subsec:heating}

In \S\ref{subsec:spectra}, we attributed the steepening of the magnetic spectrum in the sub-ion-Larmor range to the energization of ion particles through stochastic and cyclotron heating. Here, we provide evidence for this interpretation, using data taken from both the $\beta_{\rm i0}=1/9$ and $0.3$ simulations. In particular, we examine the (gyrotropized) ion distribution function $f(w_\parallel,w_\perp)$ alongside direct measures of $\rmd Q_\perp/\rmd w_\perp$ and $\rmd Q_\perp/\rmd \log k_\perp$ from these simulations, which in turn enable the evaluation of $D_{\perp\perp}^E$ via Equation (\ref{eq:dQprpdvprp_def}).
These quantities are then compared to the theoretical predictions presented in \S\ref{sec:theory}. Namely, the actual $\delta\Phi_\lambda$ fluctuation spectrum obtained from 80 (50) snapshots of the $\beta_{\rm i0}=1/9~(0.3)$ simulation during its quasi-steady state is employed in the expression for the diffusion coefficient (Equation \ref{eq:Dperp_Phi_general_xi_EXPcorr}) and the associated differential heating (Equation \ref{eq:dQprpdvprp_def}), including the exponential correction; these quantities are then time-averaged. 
At the same time, we employ an analogous procedure that considers only the $\delta u_{\perp,\lambda}$ or $\delta B_{\|,\lambda}$ fluctuations' spectrum in the approximate expression for $D_{\perp\perp}^E$ (Equation \ref{eq:Dperp_Phi_asymptotic}, including the exponential suppression term); this allows us to separate out the MHD and ``kinetic'' (non-MHD) contributions to the diffusion coefficient and to the associated differential heating (Equation \ref{eq:dQprpdvprp_def}).

\subsubsection{Ion-heating diagnostics}\label{subsubsec:heating_diagnostics}

To obtain the differential heating rate in the simulations, the following procedures have been implemented in the \pegpp code~\citep[see also][]{ArzamasskiyAPJ2019}.
At a given time, the differential rate of perpendicular heating in velocity space is computed as the sum of the instantaneous rate of work done by the electric field on each particle $p$. Namely, we compute
\begin{equation}\label{eq:Qperp_sim_def}
\widetilde{Q}_\perp=\frac{\partial^2\,Q_\perp}{\partial w_\|\partial w_\perp} \doteq\sum_p \bb{E}_{\perp p}\bcdot\bb{w}_{\perp p}
\end{equation}
and 
\begin{equation}\label{eq:Qpar_sim_def}
\widetilde{Q}_\|=\frac{\partial^2\,Q_\|}{\partial w_\|\partial w_\perp} \doteq\sum_p E_{\| p}\,w_{\| p}\,,
\end{equation}
where $\bb{E}_{p}\doteq\bb{E}(\bb{x}_p)$ is the electric field at the position $\bb{x}_{p}$ of the particle $p$ with peculiar velocity $\bb{w}_{p}\doteq\bb{v}_{p}-\bb{u}_{p}$, where $\bb{u}_{p}\doteq\bb{u}(\bb{x}_{p})$ is the mean-flow velocity at the  particle's position. 
Here $\perp$ and $\|$ are defined with respect to the actual magnetic-field direction at location $\bb{x}_{p}$: $\bb{w}_{p}=w_{\| p}\bb{b}_{p} + \bb{w}_{\perp p}$ and $\bb{E}_{p}=E_{\| p}\bb{b}_{p} + \bb{E}_{\perp p}$, with $\bb{b}_{p}\doteq\bb{B}(\bb{x}_{p})/|\bb{B}(\bb{x}_{p})|$ being the local magnetic-field unit vector.
Each of the above quantities are then binned in a two-dimensional ($w_\|,w_\perp$) space, so that they are a function of the {\em gyrotropic} (peculiar) velocity space: $\widetilde{Q}_\perp(w_\|,w_\perp)$ and $\widetilde{Q}_\|(w_\|,w_\perp)$. 
The total perpendicular or parallel heating rate is obtained as their integrals over the whole $(w_\|,w_\perp)$-space. (Thus, for instance, the one-dimensional $w_\|$-integral of $\partial^2 Q_\perp/\partial w_\|\partial w_\perp$ provides $\rmd Q_\perp/\rmd w_\perp$.)
To obtain the differential rate of heating in wavenumber space, e.g., $\rmd Q_\perp/\rmd \log k_\perp$, the electric field is Fourier-transformed and then evaluated in different log-spaced $k_\perp\doteq(k_x^2+k_y^2)^{1/2}$ bins, $\bb{E}_\perp(k_{\perp,\mathrm{bin}})$, which are then used to compute the associated rate of work on all of the simulation particles.
(In this case, the rate of work is integrated over the whole $\bb{w}$-space during run time, so that the simulation output is a function of the $k_\perp$-bins only; an updated version of this diagnostic that outputs the heating rate in the whole three-dimensional $(w_\|,w_\perp,k_\perp)$ space is currently under development.)
In the following analysis, all of the above quantities are time-averaged over the quasi-steady state (hereafter denoted by $\langle\,\cdot\,\rangle$).

\subsubsection{Free parameters in theoretical predictions}\label{subsubsec:free_parameters}

When the theoretical predictions presented in \S\ref{sec:theory} are computed from the actual fluctuation spectra obtained from the simulations, the theory has essentially three free parameters: (i) a normalization constant in Equation \eqref{eq:Dperp_Phi_general_xi_EXPcorr}, (ii) an order-unity constant $\kappa_0$ that specifies the ``resonance-like condition'' $k_\perp w_\perp/\Omega_{\rm i0}=\kappa_0$ that is used to transform the fluctuations' spectra from wavenumber to perpendicular-velocity space, {\em viz.}~$\delta\Phi(w_\perp) \longleftrightarrow \delta\Phi(k_\perp)|_{k_\perp=\kappa_0\Omega_{\rm i0}/w_\perp}$, and (iii) the constant $c_*$ in the exponential suppression factor. The constant in (i) is determined by normalizing the perpendicular-energy diffusion coefficient obtained from the $\delta\Phi_{\rm tot}$ fluctuations' spectra (Equation~\ref{eq:Dperp_Phi_general_xi_EXPcorr}) to the $D_{\perp\perp}^E$ directly obtained from the simulation at a single velocity point in the $w_\perp\leq v_{\rm th,i0}$ range (the exact point used in the following being $w_\perp/v_{\rm th,i0}=0.8$, but we verified that using any value in the range $0.5\lesssim w_\perp/v_{\rm th,i0}\lesssim1$ did not qualitatively change the results). This very same normalization constant is then used consistently for all the theoretical curves, i.e., ${\rm d}Q_\perp/{\rm d}w_\perp$ and ${\rm d}Q_\perp/{\rm d}\log k_\perp$, as well as for the theoretical predictions obtained via the different contributions to the total potential ({\em viz.}, $\delta\Phi_{\rm mhd}$ and $\delta\Phi_{\rm kin}$). Concerning the value of $\kappa_0$ and $c_*$, we show the plots when ($\kappa_0$, $c_*$) = ($1.1$, $0.09$) are adopted for the $\beta_{\rm i0}=0.3$ simulation and ($\kappa_0$, $c_*$) = ($1.25$, $0.05$) are used in the $\beta_{\rm i0}=1/9$ case. These values seem to ``best fit'' the simulations' results. The difference in the two values of $\kappa_0$ accounts somewhat for the different duration of the quasi-steady turbulent stage in the two simulations, and thus of the consequent total absolute heating of the ions during the runs (i.e., how $\rho_{\rm th,i}$ changes in the longer $\beta_{\rm i0}=0.3$ simulation). Nevertheless, we have verified that as long as it is in the range $0.9\lesssim\kappa_0\lesssim1.4$ the results do not change qualitatively. 
For what concerns the difference in the two values of $c_*$, we interpret it as the result of a different level of intermittency within the two runs (being larger at lower $\beta$). 
We have verified that, when varying $\kappa_0$, $c_*$ can also be slightly adjusted without qualitatively changing the results: values in the range $0.045\lesssim c_*\lesssim0.055$ are allowed at $\beta_{\rm i0}=1/9$, while the same holds for a range of values $0.08\lesssim c_*\lesssim0.11$ at $\beta_{\rm i0}=0.3$ (this case being less well constrained due to the higher errors associated to the ${\rm d}Q_\perp/{\rm d}\log k_\perp$ diagnostics around $k_\perp\rho_{\rm th,i}\lesssim1$; see Figure \ref{fig:Qprp_vs_kprp} and footnote \ref{footnote:beta03_dQdlogk_diagnostics}).

\subsubsection{Velocity-space dependence of ion heating}\label{subsubsec:Qprp_vs_wprp}

\begin{figure*}
    \centering
    \includegraphics[width=\textwidth]{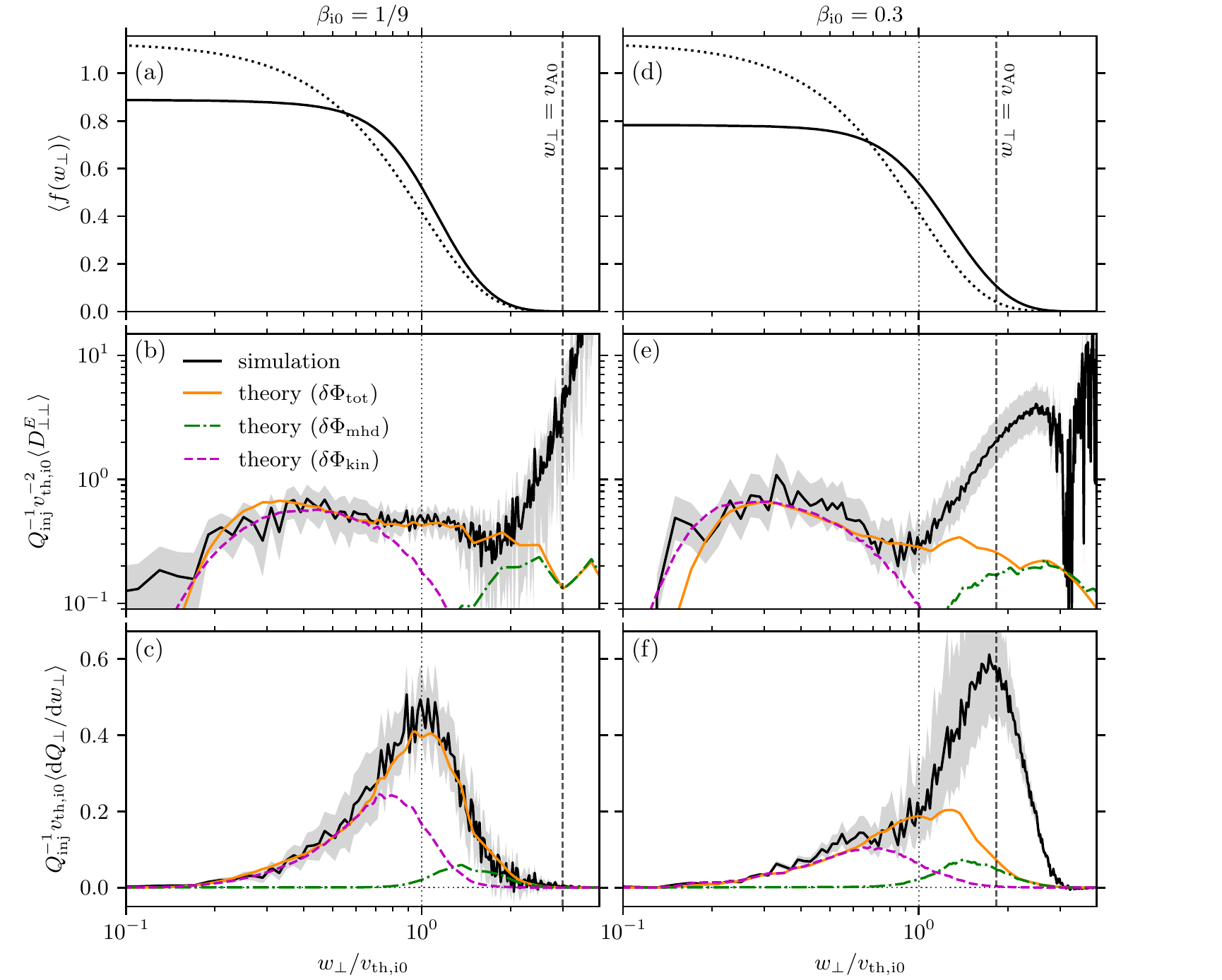}
    \caption{Left column: Comparison between the stochastic-heating theory presented in \S\ref{sec:theory} and $\beta_{\mathrm{i}0}=1/9$ simulation results versus $w_\perp/v_{\mathrm{th,i}0}$. (a) Perpendicular distribution function averaged over the quasi-steady turbulent state, $\langle f(w_\perp)\rangle$ (solid line; dotted line shows the initial Maxwellian distribution for reference). (b) Averaged perpendicular-energy diffusion coefficient, $\langle D_{\perp\perp}^E\rangle$, from simulation (black solid line) and from theory (using Equation~\eqref{eq:Dperp_Phi_asymptotic} with the exponential suppression factor) when the full potential ($\delta\Phi_{\rm tot}$; continuous orange line) or only its ideal ($\delta\Phi_{\rm mhd}$; green dot-dashed line) or non-ideal ($\delta\Phi_{\rm kin}$; purple dashed line) contribution is used. (c) Averaged differential perpendicular heating, $\langle\mathrm{d}Q_\perp/\mathrm{d}w_\perp\rangle$. Right column: Same as left column, but using results from the $\beta_{\mathrm{i}0}=0.3$ simulation.
    }
    \label{fig:Diff-Qprp_vs_wprp}
\end{figure*}

We begin by examining how the ion perpendicular distribution function $\langle f(w_\perp)\rangle$, the perpendicular-energy diffusion coefficient $\langle D_{\perp\perp}^E\rangle$, and the associated differential perpendicular heating $\langle\rmd Q_\perp/\rmd w_\perp\rangle$ behave in $w_\perp$ space. These quantities are traced by the solid black lines in Figure~\ref{fig:Diff-Qprp_vs_wprp}; results from $\beta_{\rm i0}=1/9$ ($0.3$) are in the left (right) column. These are to be compared with the theoretical predictions derived in \S\ref{sec:theory} for the diffusion and heating coefficients obtained using the spectra of the total electrostatic potential (solid orange line), of the MHD part of the potential (dash-dotted green line), and of the ``kinetic'' (i.e., non-MHD) part of the potential (dashed purple line). 

In both simulations we observe an evolution of the perpendicular distribution function, $f(w_\perp)$, from its initial Maxwellian (dotted black lines) towards a broader shape with a flat-top core (solid black lines). This evolution is the consequence of the heating mechanisms operating in the turbulence. In particular, we attribute the development of a flattened core to stochastic heating, following  \citet{KleinChandranAPJ2016}. This interpretation is supported by the two lower panels of this figure, in which both the diffusion coefficient $D_{\perp\perp}^E$ and the differential heating $\rmd Q_\perp/\rmd w_\perp$ are fit reasonably well by the theoretical curve for $w_\perp\lesssim v_{\rm th,i0}$, i.e., where the flat-top core develops.\footnote{The differential perpendicular energization ${\rm d}Q_\perp/{\rm d}w_\perp$, as measured in our simulations, exhibits some (sub-dominant) cooling effects at $w_\perp/v_{\rm th,i0}\gtrsim 2$. Because these cooling features are also present at very early times (including the initial time, $t=0$), they are likely due to errors associated with numerical noise and interpolation of the fields to the particle positions. We have modeled this cooling feature using the first few snapshots of a simulation and removed it from $\langle{\rm d}Q_\perp/{\rm d}w_\perp\rangle$ in the quasi-steady state. While we have verified that this cooling correction behaves sensibly when applied at late times (see Fig.~\ref{fig:diffusion_comparison} in Appendix~\ref{app:sec:diffusion}), one should consider the simulation curves in Figure \ref{fig:Diff-Qprp_vs_wprp} to be most reliable for $w_\perp/v_{\rm th,i0}\lesssim 2$.} From these curves, it is also evident how the relative importance of the contribution to the total stochastic ion heating from different fluctuations changes with the plasma beta: as $\beta_{\rm i0}$ decreases, the non-ideal contribution to the electrostatic potential responsible for the stochastic heating of the ions, $\delta\Phi_{\lambda,{\rm kin}}\propto(1+\tau_\perp)^{-1}\delta B_{\|,\lambda}$, becomes progressively more important than its ideal counterpart, $\delta\Phi_{\lambda,{\rm mhd}}\propto\lambda\,\delta u_{\perp,\lambda}$ (cf.~Equations~(\ref{eq:elec_potential_kaw})--(\ref{eq:Dperp_Phi_asymptotic}) and the accompanying discussion). This is highlighted by plotting explicitly the theoretical perpendicular diffusion coefficient (and the associated differential perpendicular heating) when only the ideal ($\delta\Phi_{\rm mhd}$; green dot-dashed line) or the non-ideal ($\delta\Phi_{\rm kin}$; purple dashed line) contributions to the total electrostatic potential ($\delta\Phi_{\rm tot}$; continuous orange line) are used.\footnote{Note that, while $\delta\Phi_{\rm tot,\lambda}$ is obtained as the potential part of the actual $\delta E_{\perp,\lambda}$ fluctuations, the two components $\delta\Phi_{\rm mhd,\lambda}$ and $\delta\Phi_{\rm kin,\lambda}$ are obtained via the approximate formulas using $\delta u_{\rm \perp i,\lambda}$ and $\delta B_{\|,\lambda}$, respectively (i.e., where approximate perpendicular pressure balance has been used to rewrite $\delta n_\lambda$ fluctuations in terms of $\delta B_{\|,\lambda}$, and neglecting the anisotropy correction $\Theta_\lambda\delta B_{\perp,\lambda}$; see Equation \ref{eq:elec_potential_kaw}). For this reason, the curves obtained via the approximate formulas do not exactly overlap with the one obtained using the actual $\delta\Phi_{\rm tot}$, especially at small $w_\perp$ (corresponding to small-scale wavelengths $\lambda$) where different fields (namely, $\delta n$ and $\delta B_\|$) are affected differently by numerical filters in the code.} However, Figure~\ref{fig:Diff-Qprp_vs_wprp} also shows that theoretical curves fit neither the diffusion coefficient $D_{\perp\perp}^E$ nor the differential heating $\rmd Q_\perp/\rmd w_\perp$ over the full range of $w_\perp$. This can be understood by considering the fact that (i) stochastic heating is not the only mechanism involved in the heating of ions in our simulation, and (ii) the differential heating in Figure~\ref{fig:Diff-Qprp_vs_wprp} is the result of an integration over $w_\|$ of a more structured $\widetilde{Q}_\perp(w_\|,w_\perp)$. A discussion of heating signatures within the two-dimensional $(w_\|,w_\perp)$ space is provided in \S\ref{subsec:heating_2V-space}.

\begin{figure}
    \centering
   \includegraphics[width=\columnwidth]{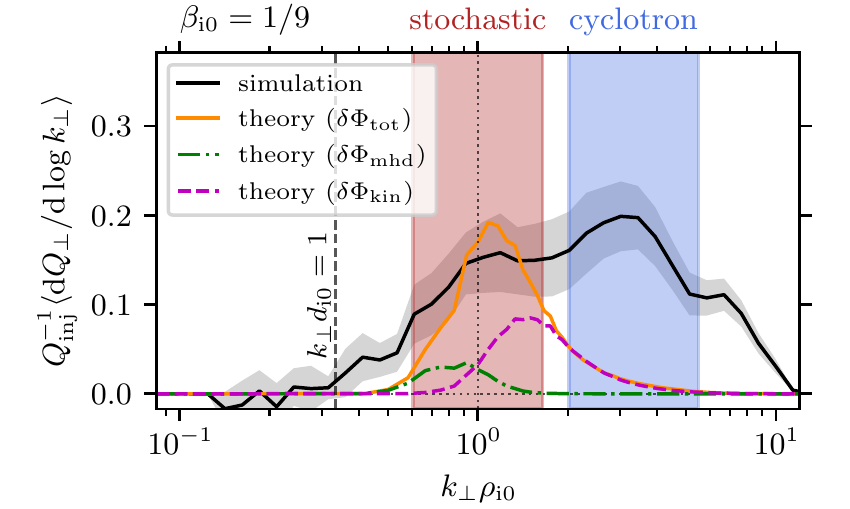}\\   \vspace{1ex}
    \includegraphics[width=\columnwidth]{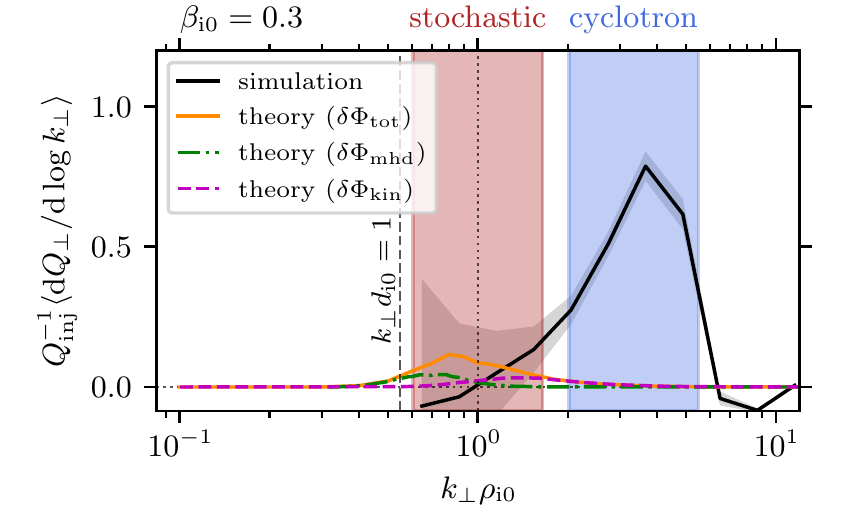}
    \caption{Top panel: Differential perpendicular energization averaged over the quasi-steady turbulent state, $\langle\mathrm{d}Q_\perp/\mathrm{d}\log k_\perp\rangle$, versus $k_\perp\rho_{\mathrm{i}0}$ in the $\beta_{\mathrm{i}0}=1/9$ simulation. The numerical result (black solid line) is compared with the theoretical prediction using the spectrum of the total electrostatic potential fluctuations in Equation \eqref{eq:elec_potential_kaw}, $\delta\Phi_{\lambda,{\rm tot}}$ (orange solid line), and the approximations considering only the $\delta\Phi_{\lambda,{\rm mhd}}$ spectrum (green dashed line) or the $\delta\Phi_{\lambda,{\rm kin}}$ spectrum (purple dashed line); the exponential correction is included, with $c_*=0.05$. The plots are obtained using the relation $k_\perp w_\perp/\Omega_{\mathrm{i}0}=\kappa_0$ with $\kappa_0=1.25$ to best fit the simulation's results in both the velocity and wavenumber spaces. The light-red (light-blue) shaded region shows the $k_\perp$ range where stochastic (ion-cyclotron) heating is considered to be important.
    Bottom panel: Same as top panel, but for the $\beta_{\mathrm{i}0}=0.3$ simulation. Here, $c_*=0.09$ and $\kappa_0=1.1$ have been adopted.}
    \label{fig:Qprp_vs_kprp}
\end{figure}

\subsubsection{Fourier-space dependence of ion heating}\label{subsubsec:Qprp_vs_kprp}

Figure~\ref{fig:Qprp_vs_kprp} displays the complementary diagnostic, the (averaged) differential heating in wavenumber space $\langle\rmd Q_\perp/\rmd\log k_\perp\rangle$, measured in the $\beta_{\rm i0}=1/9$ run (upper panel; black solid line) and the $\beta_{\rm i0} = 0.3$ run (bottom panel; black solid line). Overlaid are the theoretical curves corresponding to Equation~\eqref{eq:dQprpdvprp_def} using the total fluctuating potential (orange solid line), the ``MHD'' part of the potential (green dot-dashed line), and the ``kinetic'' part of the potential (purple dashed line).\footnote{To obtain the theoretical predictions plotted in Figure~\ref{fig:Qprp_vs_kprp}, the  theoretical lines of ${\rm d}Q_\perp/{\rm d}w_\perp$ corresponding to Equation~\eqref{eq:dQprpdvprp_def}, which are plotted in Figure~\ref{fig:Diff-Qprp_vs_wprp}, have been interpolated into $\log k_\perp$ space. This procedure also takes into account the logarithmic spacing of the volume in passing from ${\rm d}w_\perp$ to ${\rm d}\log k_\perp$, i.e., that ${\rm d}Q_\perp/{\rm d}\log k_\perp = (\kappa_0\Omega_{\rm i0}/k_\perp)\big[{\rm d}Q_\perp/{\rm d}w_\perp\big]_{w_\perp=\kappa_0\Omega_{\rm i0}/k_\perp}$.}

At $\beta_{\rm i0}=1/9$, the differential heating clearly exhibits two distinct peaks in the perpendicular-wavenumber space: one at $k_\perp\rho_{\rm i0}\approx1$, and a second one at $k_\perp\rho_{\rm i0}\approx3$. We interpret the first peak as the result of stochastic ion heating, consistent with the theoretical curves obtained when the actual fluctuations' spectra are employed in the expressions derived in \S\ref{sec:theory}. The second peak at $k_\perp\rho_\mathrm{i0}\approx 3$ is interpreted as being due to ion-cyclotron heating associated with the $n=1$ cyclotron resonance, consistent with the fact that the frequency of the fluctuations reaches $\omega\approx\Omega_{\rm i0}$ at such a value of $k_\perp\rho_{\rm i0}$ (see Figure~\ref{fig:spectra} and accompanying discussion). An additional (minor) contribution to the total ion heating can be seen at $k_\perp\rho_{\rm i0}\gtrsim6$, likely associated with the $n>1$ cyclotron resonances discussed in \S\ref{subsec:spectra}). These two mechanisms, {\em stochastic} and {\em ion-cyclotron} heating, contribute roughly equally to the overall perpendicular heating of the ions at $\beta_{\rm i0}=1/9$: $Q_\perp^{\rm stoch}/Q_\perp^{\rm tot}\approx Q_\perp^{\rm cycl}/Q_\perp^{\rm tot}\approx50$\%.

The overall perpendicular ion heating at $\beta_{\rm i0}=0.3$ (Figure \ref{fig:Qprp_vs_kprp}, bottom) is dominated by scales at which we expect ion-cyclotron heating to be important; stochastic heating accounts for at most a quarter of the total heating: $Q_\perp^{\rm cycl}/Q_\perp^{\rm tot}\gtrsim75$\% and $Q_\perp^{\rm stoch}/Q_\perp^{\rm tot}\lesssim25$\%.\footnote{The older $\beta_{\rm i0}=0.3$ simulation employed a heating diagnostic that used the total particle velocity $\bb{v}_p$ in Equations (\ref{eq:Qperp_sim_def}) and (\ref{eq:Qpar_sim_def}) rather than its peculiar velocity $\bb{w}_p$ (as in the version of the diagnostic employed in the new $\beta_{\rm i0}=1/9$ run). Also, the $k_\perp$ resolution used to compute this diagnostic was lower in the $\beta_{\rm i0}=0.3$ run (12 bins) than for $\beta_{\rm i0}=1/9$ (40 bins). As a result, the error bars on the heating at $k_\perp\rho_{\rm i0}\lesssim 1$ are much larger in the $\beta_{\rm i0}=0.3$ run.\label{footnote:beta03_dQdlogk_diagnostics}}

An important trend that arises from the above analysis is that (i) stochastic ion heating should become progressively more important than ion-cyclotron heating as the plasma $\beta$ decreases, and (ii) this result is mainly due to contributions from the non-ideal electric field (and associated potential, $\delta\Phi_{\rm kin}$) arising from the Hall and thermo-electric effects in Equation (\ref{eq:generalized_Ohm}). In fact, while the ideal contribution to the stochastic heating from $\delta\Phi_{\rm mhd}$ is nearly constant when passing from $\beta_{\rm i0}=0.3$ to $\beta_{\rm i0}=1/9$, the heating associated with $\delta\Phi_{\rm kin}$ nearly doubles in its contribution. This in turn lowers the amount of the fluctuations' energy that is available when the ion-cyclotron frequency is reached in the cascade, consequently diminishing the contribution of the ion-cyclotron mechanism to the overall ions' perpendicular heating.

\subsection{Intermittency contributions to stochastic heating}\label{subsec:tracked_particle}

\begin{figure}
    \centering
    \includegraphics[width=\columnwidth]{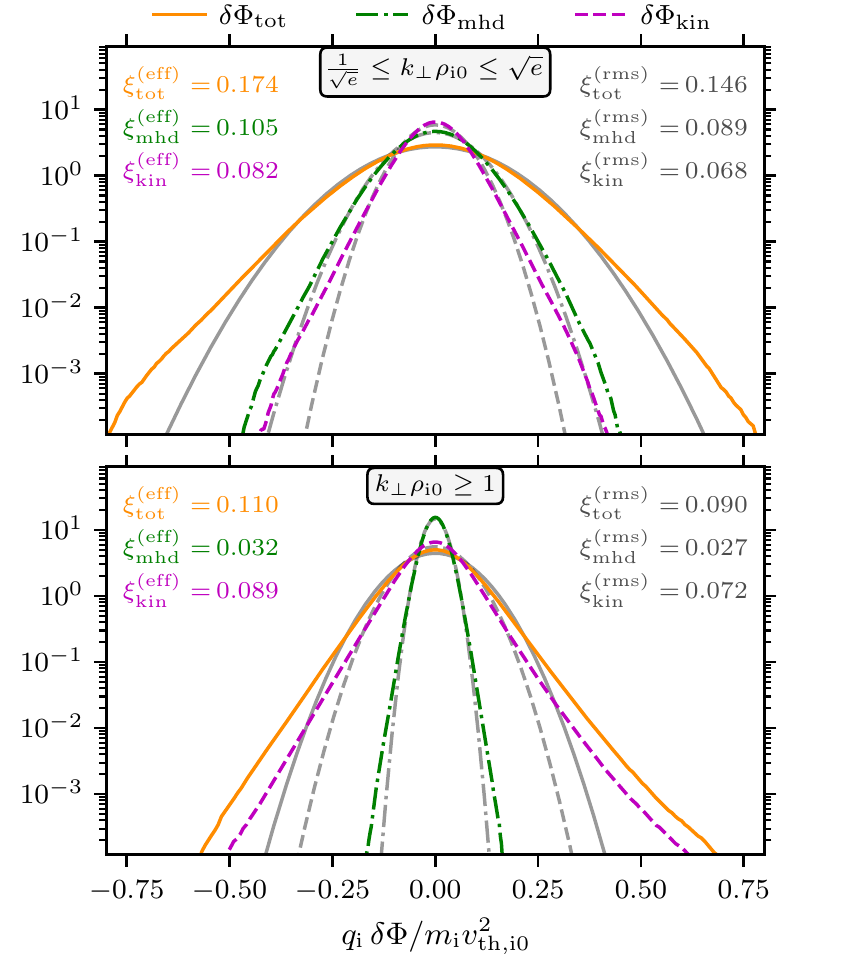}%
    \caption{PDF of (normalized) potential fluctuations, $q_{\rm i}\delta\Phi/m_{\rm i}v_{\rm th,i0}^2$, from our $\beta_{\rm i0}=1/9$ simulation, in the range of scales in which stochastic heating is considered to be the dominant ion-heating mechanism, $1/\sqrt{\rme}\leq k_\perp\rho_{\rm i0}\leq\sqrt{\rme}$ (upper panel), and at all sub-ion-gyroradius scales, $k_\perp\rho_{\rm i0}\geq1$ (lower panel). Statistics of both the total potential, $\delta\Phi_{\rm tot}$ (orange solid), and its ideal and non-ideal parts, $\delta\Phi_{\rm mhd}$ (green dot-dashed) and $\delta\Phi_{\rm kin}$ (purple dashed) respectively, are reported. Equivalent Gaussian statistics are also drawn as grey lines (with corresponding line style). Both the ``effective'' and rms value of the stochasticity parameter (computed using the actual PDF of the fluctuations) is reported in each plot as $\xi^{\rm(eff)}$ and $\xi^{\rm(rms)}$, respectively. In the range $1/\sqrt{\rme}\leq k_\perp\rho_{\rm i0}\leq\sqrt{\rme}$ (upper panel), $q_{\rm i}\delta\Phi/m_{\rm i}v_{\rm th,i0}^2$  corresponds roughly to the thermal stochasticity parameter, $\xi_{\rm th}$, which estimates the overall efficiency of stochastic heating (see \S\ref{subsec:theory_revisited}).}
    \label{fig:PDF_Phi}
\end{figure}
\begin{figure}
    \centering
    \includegraphics[width=\columnwidth]{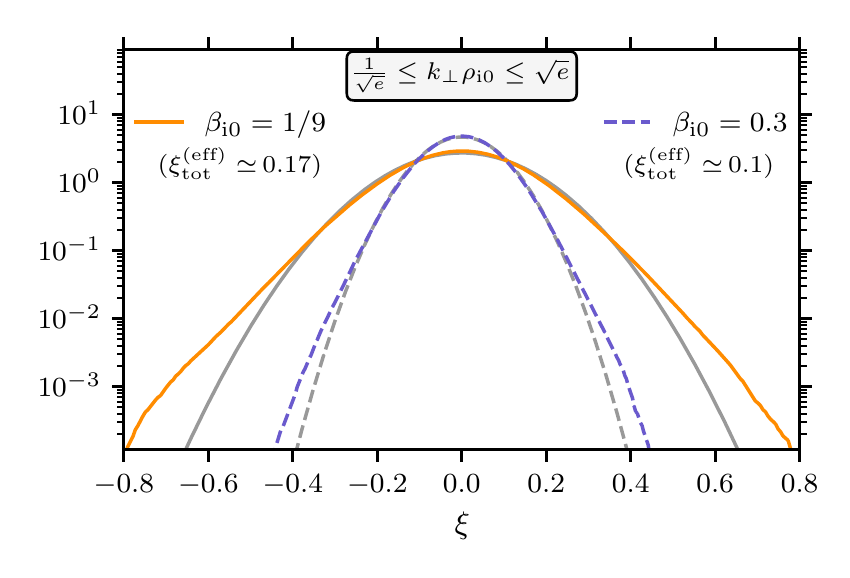}\\
    \includegraphics[width=\columnwidth]{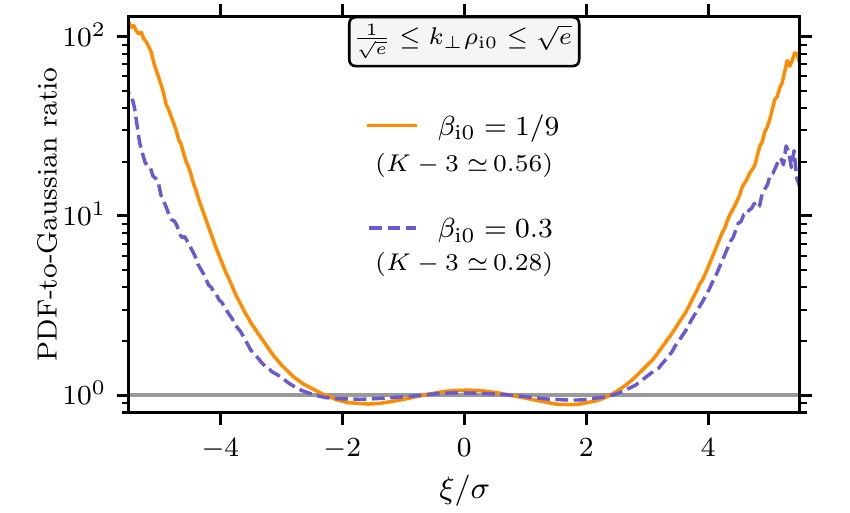}%
    \caption{Top: Comparison between PDF of (normalized) total potential fluctuations, $\xi\doteq q_{\rm i}\delta\Phi_{\rm tot}/m_{\rm i}v_{\rm th,i0}^2$, from the $\beta_{\rm i0}=1/9$ (orange solid line) and $\beta_{\rm i0}=0.3$ (violet dashed line) simulations in the range of scales $1/\sqrt{\rme}\leq k_\perp\rho_{\rm i0}\leq\sqrt{\rme}$. (The corresponding value of $\xi_{\rm tot}^{\rm(eff)}$ is also reported, below each simulation label). Grey lines (with corresponding line style) represent the equivalent Gaussian distribution characterized by the same standard deviation $\sigma$ as the actual PDF. Bottom: Comparison of deviation from Gaussian statistics for the potential fluctuations (still in the range $1/\sqrt{\rme}\leq k_\perp\rho_{\rm i0}\leq\sqrt{\rme}$) in the $\beta_{\rm i0}=1/9$ (orange solid) and $\beta_{\rm i0}=0.3$ (violet dashed) simulations. This deviation is quantified both by the ratio of the actual PDF and the equivalent Gaussian versus $\xi/\sigma$ (colored lines), and by the ``excess kurtosis'', $K-3$ (reported in the plot, below each simulation label;  $K\doteq\langle\xi^4\rangle/\langle\xi^2\rangle^2=3$ for a zero-mean Gaussian distribution).}
    \label{fig:PDF_Phi_beta-comparison}
\end{figure}

To explore the degree of intermittency of the potential fluctuations (and its effect on the stochastic heating) in the $\beta_{\rm i0}=1/9$ simulation, in Figure~\ref{fig:PDF_Phi} we report the probability density function (PDF) of the normalized total potential fluctuations, $q_{\rm i}\delta\Phi_{\rm tot}/m_{\rm i}v_{\rm th,i0}^2$ (orange solid line), and of its ideal and non-ideal parts, $q_{\rm i}\delta\Phi_{\rm mhd}/m_{\rm i}v_{\rm th,i0}^2$ (green dot-dashed line) and $q_{\rm i}\delta\Phi_{\rm kin}/m_{\rm i}v_{\rm th,i0}^2$ (purple dashed line), respectively. Equivalent Gaussian distributions are also drawn as grey lines (with the same line-style of the potential contribution to which they correspond). These PDFs are computed on two different ranges of scales: (i) $1/\sqrt{\rme}\leq k_\perp\rho_{\rm i0}\leq\sqrt{\rme}$ (upper panel), corresponding to the range where stochastic heating is considered to be the dominant ion-heating mechanism, and (ii) $k_\perp\rho_{\rm i0}\geq1$ (lower panel), corresponding to the entire sub-ion-gyroradius (``kinetic'') range.

From a statistical point of view, Figure~\ref{fig:PDF_Phi} clearly shows that, while the width of the overall fluctuation-amplitude distribution decreases towards smaller scales, the degree of intermittency of these fluctuations simultaneously increases. Both aspects are relevant for the enhancement of stochastic ion heating. Let us consider the range of scales reported in the upper panel in Figure~\ref{fig:PDF_Phi} ({\em viz.} $1/\sqrt{\rme}\leq k_\perp\rho_{\rm i0}\leq\sqrt{\rme}$).
In this range around $k_\perp\rho_{\rm i0}\sim1$, the quantity $q_{\rm i}\delta\Phi_{\rm tot}/m_{\rm i}v_{\rm th,i0}^2$ corresponds to (a generalized version of) the stochasticity parameter that has been previously used to estimate the efficiency of stochastic heating~\citep[e.g.,][]{XiaAPJ2013,VasquezAPJ2015,MartinovicAPJS2020}. First, one notices that the distribution of fluctuations' amplitudes itself is relatively broad in this simulation, even for an equivalent-Gaussian distribution: this implies that, even without taking into account intermittency, gyro-scale fluctuations are not negligibly small. This is further quantified by computing both the rms stochastic-heating parameter, $\xi^{\rm(rms)}$, and an effective value, $\xi^{\rm(eff)}$, that takes into account the non-Gaussian nature of the actual fluctuations' PDFs.\footnote{Because the heating is proportional to $|q_{\rm i}\delta\Phi/m_{\rm i}v_{\rm th,i0}^2|^3\approx|\xi|^3$, we define this effective parameter by $\xi^{\rm(eff)}=\big[\int{\rm d}\xi\,|\xi|^3\,{\cal P}(\xi) \big]^{1/3}$, where ${\cal P}$ is the actual PDF of $q_{\rm i}\delta\Phi/m_{\rm i}v_{\rm th,i0}^2$.} These values are reported in each panel for the different scale ranges considered. Even in its rms version, within both scale ranges the stochasticity parameter is large enough ($\xi\gtrsim0.1$) that the overall effect of an exponential suppression term in \eqref{eq:Dperp_Phi_general_xi_EXPcorr} should be small if $c_*\approx0.01$--$0.1$. Second, intermittency does enhance the effective stochasticity parameter (and the associated heating). In fact, in the range of scales around $k_\perp\rho_{\rm i0}\sim1$ (upper panel of Figure~\ref{fig:PDF_Phi}), intermittency increases $\xi^{\rm(rms)}$ by ${\approx}19$\%.
This effect is more important when the whole sub-ion range of scales is considered, $k_\perp\rho_{\rm i0}\geq1$ (lower panel in Figure~\ref{fig:PDF_Phi}): over this range of scales, $\xi^{\rm (eff)}$ is increased beyond its equivalent-rms value $\xi^{\rm(rms)}$ by ${\approx}22$\% (although the absolute values of $\xi$ in this range are indeed smaller than the corresponding values in the range around $k_\perp\rho_{\rm i0}\sim1$).

The degree of intermittency also appears to depend on $\beta_{\rm i}$. In the top panel of Figure~\ref{fig:PDF_Phi_beta-comparison}, we report a comparison between the PDFs of the normalized total potential fluctuations, $\xi\doteq q_{\rm i}\delta\Phi_{\rm tot}/m_{\rm i}v_{\rm th,i0}^2$, around $k_\perp\rho_{\rm th,i}\sim1$ in the $\beta_{\rm i0}=1/9$ simulation (orange solid line) and in the $\beta_{\rm i0}=0.3$ simulation (violet dashed line). 
It is evident that the fluctuations' distribution broadens significantly at lower $\beta_{\rm i0}$, passing from $\xi_{\rm tot}^{\rm(eff)}\approx0.1$ at $\beta_{\rm i0}=0.3$ to  $\xi_{\rm tot}^{\rm(eff)}\approx0.17$ at $\beta_{\rm i0}=1/9$. This demonstrates that stochastic heating is enhanced as the plasma $\beta$ decreases, as expected. But we also find that the level of intermittency increases at lower $\beta$. In the bottom panel of Figure~\ref{fig:PDF_Phi_beta-comparison}, we report the ratio between the actual PDF of $\xi$ and an equivalent-width Gaussian distribution characterized by the same standard deviation $\sigma$ of the actual PDF (because $\sigma$ depends on $\beta_{\rm i0}$, the ratio is plotted versus $\xi/\sigma$ for the comparison to be meaningful). 
This PDF-to-Gaussian ratio exhibits larger deviations from unity at $\beta_{\rm i0}=1/9$ (orange solid line) than it does at $\beta_{\rm i0}=0.3$, a feature we further quantify by calculating the so-called ``excess kurtosis'', $K-3$ (with the kurtosis defined by $K\doteq\langle\xi^4\rangle/\langle\xi^2\rangle^2$; $K=3$ for a Gaussian distribution with zero mean). This quantity doubles passing from $\beta_{\rm i0}=0.3$ (for which $K-3\approx0.28$) to $\beta_{\rm i0}=1/9$ (being $K-3\approx0.56$).
We interpret this enhanced intermittency as being responsible for decreasing the effective value of $c_*$ needed to fit our simulation results at different $\beta_{\rm i0}$. Further numerical and observational studies are needed to determine the exact dependence of $c_*$ on the plasma parameters.

\begin{figure*}
    \centering
    \includegraphics[width=\textwidth]{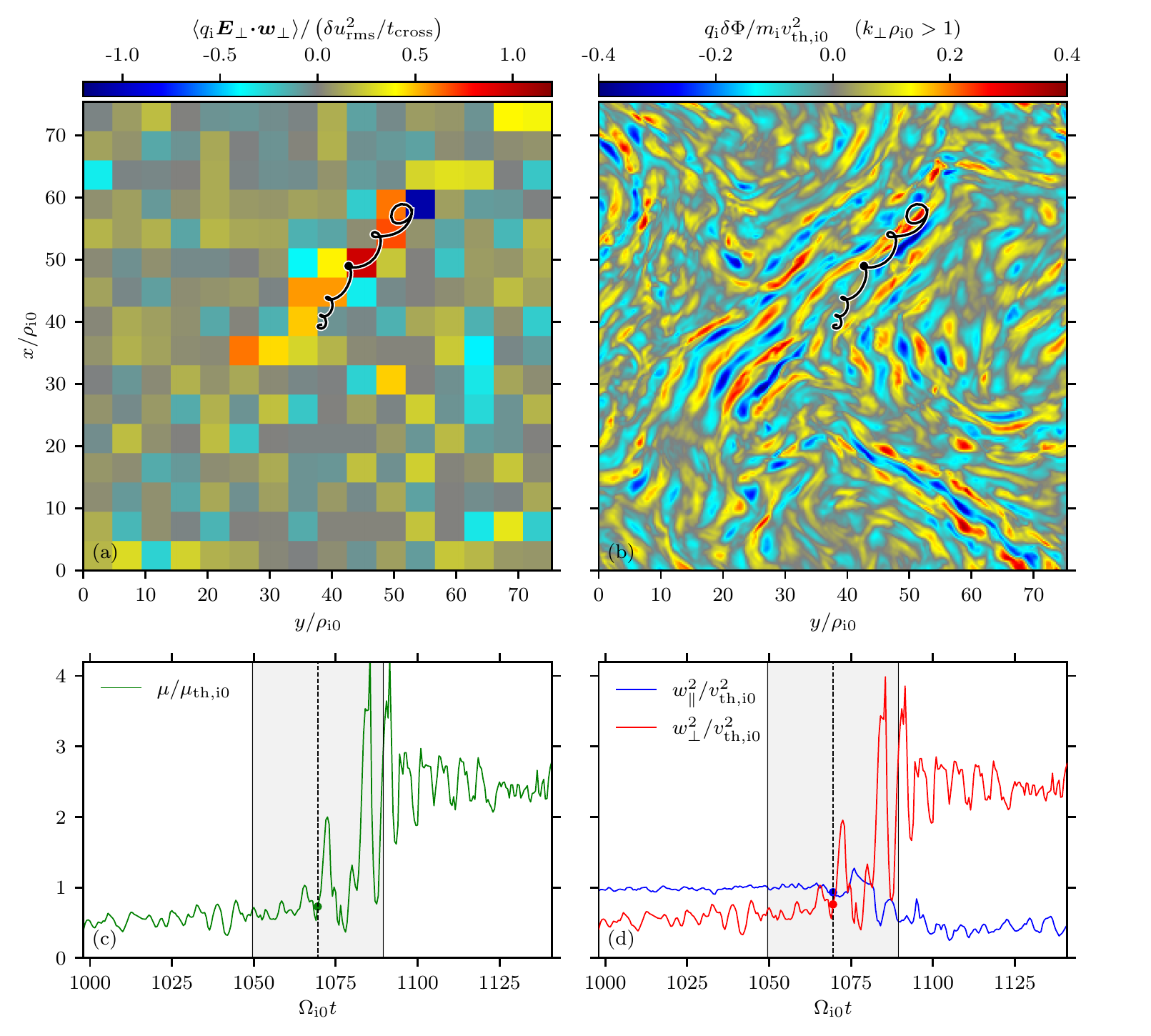}
    \caption{Example of a simulation particle undergoing stochastic heating in the $\beta_{\rm i0} = 1/9$ simulation. (top) Snapshots of ion energization averaged over $18^3$ cells (left) and small-scale ($k_\perp \rho_{\rm i0} > 1$) potential fluctuations (right) in a plane perpendicular to the guide field. Black line shows a trajectory of the particle located in this plane. This particle starts in a region with small potential fluctuations, and moves through a localized region with large $\delta \Phi$. (bottom) Evolution of the particle's magnetic moment (left), along with parallel and perpendicular energies (right). The time over which particle trajectory is plotted in the upper panels is indicated by the gray shaded region. As the particle moves through strong potential fluctuations, it undergoes non-adiabatic perpendicular heating, which changes the particle's energy by a factor of a few over a timescale of several orbits.}
    \label{fig:track}
\end{figure*}

To further illustrate the partially intermittent nature of the stochastic ion heating in the $\beta_{\rm i0}=1/9$ run, we show the evolution of one the simulation particles in Figure \ref{fig:track}. This particle was specifically chosen because it increased its energy significantly over a short period of time by interacting with an intense, spatially and temporally localized potential fluctuation.

The left upper panel (Figure~\ref{fig:track}(a)) shows the perpendicular ion energization $\langle q_{\rm i} \bb{E}_\perp \bcdot \bb{w}_\perp \rangle$ in the perpendicular (to the guide field) plane through which the tracked particle passed at that moment. The energization is averaged over multiple cells in the simulation (a volume of $18^3$ cells) to reduce the noise; it is normalized to $\delta u_{\rm rms}^2/t_{\rm cross}$, which serves as a proxy for the cascade rate. It is clear that the majority of the perpendicular energization happens in a spatially localized region. In the right upper panel (Figure~\ref{fig:track}(b)) we show the (normalized) potential fluctuations, $q_{\rm i}\delta\Phi/m_{\rm i}v_{\rm th,i}^2$, in the same plane represented in panel (a); in this case, $\delta\Phi$ has been filtered to select only those modes satisfying $k_\perp \rho_{\rm i0} > 1$. Comparing this contour plot with the one in panel (a), one can see a clear correlation between the region in which the amplitude of the potential fluctuations is larger and where most of the ion energization  occurs. The majority of the energization happens in the region in which the Larmor-scale potential fluctuations are comparable to the thermal kinetic energy of typical particle, $q_{\rm i}\delta \Phi \sim m_{\rm i} v_{\rm th,i0}^2$ (i.e., $\xi\sim1$). As discussed earlier, the reason why such potential fluctuations can be so large, even though $q_{\rm i} \delta\Phi_{\rm rms}|_{k_\perp\rho_{\rm i0}>1} \sim 0.1m_{\rm i} v^2_{{\rm th,i0}}$, is because the turbulence is intermittent (cf.~lower panel of Figure~\ref{fig:PDF_Phi}).
This picture is supported by solar-wind measurements, which show a clear correlation between coherent magnetic structures generated intermittently by strong turbulence and plasma (anisotropic) heating~\citep[e.g.,][]{OsmanPRL2012,GrecoSSR2018,QudsiAPJS2020}.\footnote{From \citet{ChandranAPJ2010}: ``\dots in strong AW/KAW turbulence (as opposed to randomly phased waves), a significant fraction of the cascade power may be dissipated in coherent structures in which the fluctuating fields are larger than their rms values. Proton orbits in the vicinity of such structures are more stochastic than in average regions, and thus $c_2$ may be smaller in AW/KAW turbulence than in our test-particle simulations, indicating stronger heating.''\label{footnote:chandran}}

Panels (c) and (d) of this figure show this tracked particle's magnetic moment $\mu$ (green line), normalized to its initial value $\mu_{\rm th,i0} \doteq m_{\rm i} v_{\rm th,i0}^2/2B_0$, and the particle's parallel and perpendicular thermal energies (blue and red lines, respectively), normalized to $v_{\rm th,i0}^2$, versus time. All of these quantities are approximately constant during particle gyration.\footnote{These quantities ($\mu$, $w_\parallel$, $w_\perp$) are calculated using the magnetic field interpolated to the particle position, rather than to the particle's guiding center. The difference is responsible for the small, periodic variations seen in these quantities on timescales ${\sim}2\pi/\Omega_{\rm i0}$.}  However, once the particle enters the region with strong potential fluctuations (the gray shaded region in these panels), its perpendicular energy and magnetic moment oscillate with large amplitude. After ${\approx}6$ gyrations, particle's perpendicular energy and its magnetic moment change by a factor of ${\approx}4.2$.

Figures \ref{fig:PDF_Phi}--\ref{fig:track} highlight further the importance of intermittency in reducing the effectiveness of the exponential suppression factor introduced by \citet{ChandranAPJ2010}, at least under the conditions realized in our simulations (see \S\ref{sec:theory}). 
This is because of the relatively large rms amplitude of gyro-scale potential fluctuations ($\xi_{\rm th}=q_{\rm i}\delta\Phi_{\rho_{\rm i}}/m_{\rm i}v_{\rm th,i0}^2\sim 0.1$) and the intermittent nature of those fluctuations, the latter of which causes a non-negligible fraction of heating to occur in localized regions exhibiting large potential fluctuations. 
As a result, a particle's energy often changes considerably during just a few gyrations, and their orbits become stochastic, so that exponential conservation of magnetic moment no longer holds.

As a final remark, we speculate that intermittency may allow stochastic heating to remain an important energization mechanism for low-$\beta$ turbulent systems even at scale separations much larger than what was achieved in our simulations~\citep[see, e.g.,][]{MalletJPP2019}. As the scale separation increases, $\delta \Phi_{\rm rms}|_{k_\perp \rho_{\rm i0} \sim 1}$ decreases but $\delta \Phi$ becomes localized within a smaller volume, creating larger potential drops within this volume. In other words, the trend outlined in Figure~\ref{fig:PDF_Phi} for our $\beta_{\rm i0}=1/9$ run suggests that, while the PDF of the fluctuations' amplitude at ion/sub-ion scales may become progressively narrower as the scale separation $L/\rho_{\rm i0}$ increases, the intermittency effects will become simultaneously more and more important in enhancing $\xi^{\rm(eff)}$ with respect to $\xi^{\rm(rms)}$.

\begin{figure*}
    \centering
    \includegraphics[width=\textwidth]{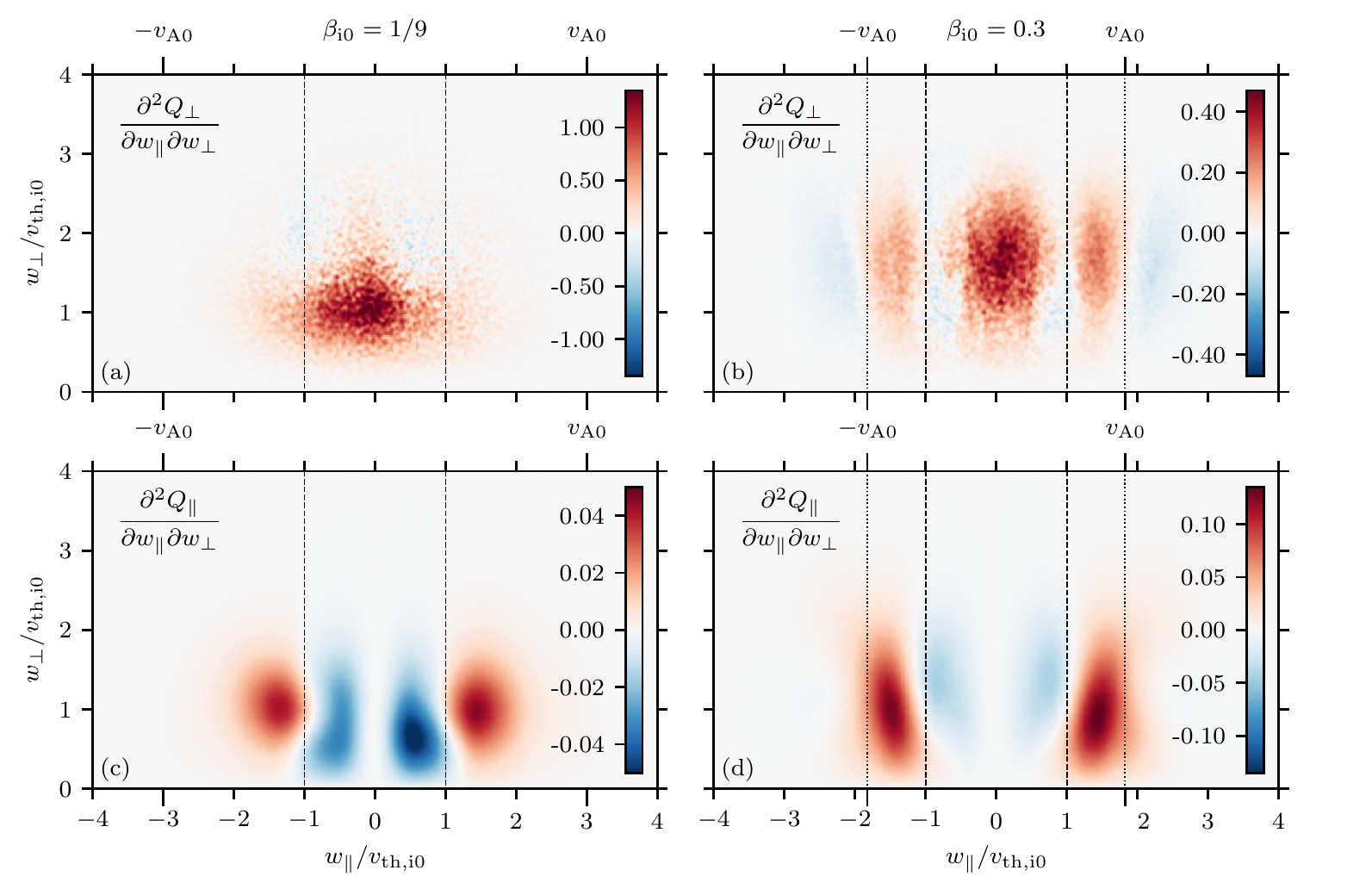}
    \caption{
    Ion energization rate as a function of parallel ($w_\parallel$) and perpendicular ($w_\perp$) velocities. Panels (a) and (c) show parallel and perpendicular energization in the $\beta_{\rm i0} = 1/9$ simulation. Panels (b) annd (d) show the same quantities for the $\beta_{\rm i0}=0.3$ simulation from \citet{ArzamasskiyAPJ2019}. The ion-thermal is marked by dashed lines; the Alfv\'en speed in the $\beta_{\rm i0} = 0.3$ run is marked by dotted lines.
}
    \label{fig:QparaQprp_vs_w_2D}
\end{figure*}

\begin{figure}
    \centering
    \includegraphics[width=\columnwidth]{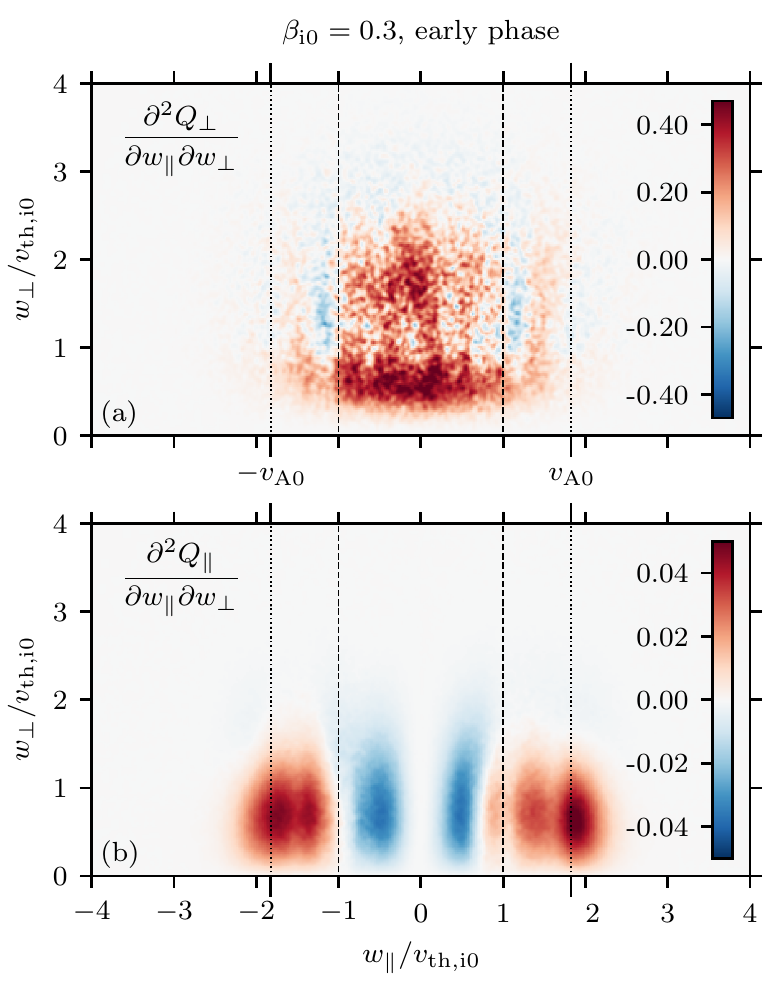}
    \caption{
    Parallel and perpendicular energization at early times in the $\beta_{\rm i0} = 0.3$ simulation from \citet{ArzamasskiyAPJ2019}, before flattening of the perpendicular-velocity core of the distribution function suppresses stochastic heating. The ion-thermal (Alfv\'{e}n) speed is marked by the dashed (dotted) lines.
}
    \label{fig:QparaQprp_vs_w_2D_beta03}
\end{figure}

\subsection{Other signatures of wave-particle interaction}\label{subsec:heating_2V-space}

The parallel and perpendicular ion-energization rates in the two-dimensional velocity space, $\widetilde{Q}_\|(w_\|,w_\perp)$ and $\widetilde{Q}_\perp(w_\|,w_\perp)$ respectively (see Equations \eqref{eq:Qperp_sim_def} and \eqref{eq:Qpar_sim_def}), can also be used to uncover the phase-space signatures of different wave-particle interactions. Their time-averaged values in the quasi-steady state, $\langle\widetilde{Q}_\|(w_\|,w_\perp)\rangle$ and $\langle\widetilde{Q}_\perp(w_\|,w_\perp)\rangle$, are reported in Figure~\ref{fig:QparaQprp_vs_w_2D} for both the $\beta_{\rm i0}=1/9$ (left column) and $\beta_{\rm i0}=0.3$ (right column) simulations. Figure~\ref{fig:QparaQprp_vs_w_2D_beta03} additionally provides this information for $\beta_{\rm i0}=0.3$ during its ``early phase'', which refers to times $t/\tau_{\rm A}\approx 3.8$--$4.4$ before the core of the perpendicular distribution function becomes appreciably flattened and stochastic heating is consequently reduced (see figure 8 of \citet{ArzamasskiyAPJ2019}).

The velocity-space patterns of $\langle\widetilde{Q}_\|(w_\|,w_\perp)\rangle$ seen in the quasi-steady state of both simulations (Figure~\ref{fig:QparaQprp_vs_w_2D}(c,d)) display the signature of collisionless damping at the Landau resonances, $w_\parallel \approx \pm v_{\rm th,i0}$ \citep[cf.][]{HowesJPP2017}. We interpret this structure as being due to the collisionless damping of slow-mode fluctuations. In the $\beta_{\rm i0}=1/9$ simulation, the amount of parallel energization associated with this Landau-resonant damping is extremely sub-dominant, contributing only ${\lesssim}2\%$ of the total ion heating rate. In the $\beta_{\rm i0}=0.3$ simulation, this percentage is ${\lesssim}10\%$. During the early phase of the $\beta_{\rm i0}=0.3$ run (Figure~\ref{fig:QparaQprp_vs_w_2D_beta03}(b)), there is an additional signature of wave-particle interaction in the vicinity of $w_\|\approx\pm\,v_{\rm A0}$. We attribute the majority of the measured increase in parallel temperature instead to a combination of transit-time damping, which is driven by $Q_\perp$ (note the vertical resonant-like red and blue ``stripes'' in Figures~\ref{fig:QparaQprp_vs_w_2D}(a,b) and \ref{fig:QparaQprp_vs_w_2D_beta03}(a)), and pitch-angle scattering of perpendicularly energized particles~\citep[as in][\S~3.2]{ArzamasskiyAPJ2019}.\footnote{\citet{IsenbergApJ2019} suggested that the perpendicularly heated ion distribution functions with $T_\perp>T_\parallel$ that are naturally generated by ion stochastic heating would be unstable to the ion-cyclotron anisotropy instability, which would then generate quasi-parallel-propagating ion-cyclotron waves and thereby scatter ions into the parallel direction. The connection between this suggestion and the pitch-angle scattering of perpendicularly energized particles measured by \citet{ArzamasskiyAPJ2019} and also seen here is not clear, for two main reasons. First, the temperature anisotropies measured in our simulations never become as large as those found in model devised by \citet{IsenbergApJ2019}; for example, $T_\perp/T_\parallel \lesssim 1.12$ in our $\beta_{\rm i0}=1/9$ simulation. Second, our steady-state perpendicular distribution functions retain flattened cores similar to those predicted by \citet{KleinChandranAPJ2016}; \citet{IsenbergApJ2019} predicted that pitch-angle scattering from unstable ion-cyclotron waves would erase this distinctive feature.}

In contrast with the $\beta_\mathrm{i0}=0.3$ case, the parallel ion distribution, $f(w_\|)$, does not develop significant non-thermal tails at $\beta_\mathrm{i0}=1/9$ (not shown). This can be explained by the inefficient Landau damping of Alfv\'{e}nic fluctuations at very low values of $\beta$: at $\beta\ll1$, the Alfv\'en speed is much larger than the ion-thermal velocity, $v_\mathrm{A}\gg v_\mathrm{th,i}$, and only the very tail of the ion distribution can effectively resonate with the phase velocity $v_\mathrm{ph}\sim v_\mathrm{A}$ of Alfv\'enic fluctuations. Since this population is energetically unimportant for the overall thermal budget of the plasma, we do not expect to find significant (parallel) heating from this process at very low $\beta$.\footnote{The same argument can also explain why, within gyrokinetic theory and simulations, the ion-to-electron heating dramatically drops at low $\beta$~\citep[e.g.,][]{HowesMNRAS2010,KawazuraPNAS2019}: because species' heating in gyrokinetics relies only on the Landau damping of the fluctuations (which can thus provide only parallel heating), Alfv\'enic turbulence will be damped inefficiently by ions as the plasma $\beta$ decreases. (The large-scale injection of compressive fluctuations, which may be collisionlessly damped even at low $\beta$, at energy levels comparable to those of the Alfv\'{e}nic fluctuations modifies this expectation; \citealt{KawazuraPRX2020}.)}

Finally, both runs display signatures that may be interpreted as the superposition of (i) stochastic heating and (ii) ion-cyclotron heating. Stochastic heating presents in both runs as a horizontal feature close to $w_\perp\sim v_{\rm th,i0}$. For $\beta_{\rm i0}=0.3$, this signature is much more pronounced during its ``early phase'' (Figure~\ref{fig:QparaQprp_vs_w_2D_beta03}(a)) than in its quasi-steady state, in which the core of the perpendicular distribution function is substantially flattened and stochastic heating is reduced. Ion-cyclotron heating, on the other hand, presents as a (fuzzy) circular halo centered around $w_\|\sim0$ and $w_\perp\gtrsim v_{\rm th,i0}$ (cf.~\citealt{KleinJPP2020}). However, the position and extension of this halo in $w_\perp$ seems to vary between $\beta_{\rm i0}=0.3$ and $\beta_{\rm i0}=1/9$; this feature is not well understood and should be investigated in future work.

\section{A comment on the interpretation of stochastic heating in spacecraft data}\label{sec:interpretation}

Before summarizing our main findings, we pause here to offer a comment on how spacecraft data might be best interpreted when looking for evidence of stochastic ion heating in the low-$\beta$ solar wind. We begin by summarizing the method adopted by \citet{Bourouaine13}, \citet{VechAPJ2017}, and  \citet{MartinovicAPJ2019,MartinovicAPJS2020}. Those authors used spacecraft-measured amplitudes of magnetic-field fluctuations near the proton gyroscale, $\delta B_{\rho_{\rm i}}$, as a proxy for the gyroscale velocity fluctuations, $\delta u_{\rm i,\rho_{\rm i}}$. The latter was then divided by the field-perpendicular proton thermal speed, $v_{\rm th,i} \doteq \sqrt{2T_{\perp,\rm i}/m_{\rm i}}$, to obtain estimates for the stochasticity parameter $\epsilon_{\rm i}$ originally introduced by \citet{ChandranAPJ2010}. (Recall footnote \ref{footnote:stochasticity}.) Specifically, they set
\begin{equation}\label{eqn:AWKAW}
    \delta u_{\rm i,\rho_{\rm i}} = \sigma \frac{\delta B_{\rho_{\rm i}}}{\sqrt{4\pi m_{\rm i} n}} ,
\end{equation}
where $\sigma$ is an order-unity constant (typically $1.19$), so that
\begin{equation}\label{eqn:epsilon}
    \epsilon_{\rm i} = \sigma \beta^{-1/2}_{\perp,\rm i} \frac{\delta B_{\rho_{\rm i}}}{B_0} ,
\end{equation}
where $B_0$ is the mean magnetic-field strength. The amplitudes of the gyroscale magnetic-field fluctuations were defined using
\begin{equation}\label{eqn:Bcentered}
    \delta B_{\rho_{\rm i}} \doteq \left[ \int^{\sqrt{\rme}\rho^{-1}_{\rm i}}_{\rho^{-1}_{\rm i}/\sqrt{\rme}} \rmd k_\perp \, E_B(k_\perp) \right]^{1/2} ,
\end{equation}
where $E_B(k_\perp)$ is the (appropriately normalized) one-dimensional magnetic energy spectrum in the plasma rest frame (obtained by applying Taylor's hypothesis to the frequency spectrum measured by the spacecraft). The amount of stochastic heating associated with these fluctuations was then inferred using
\begin{equation}\label{eq:Qprp_Ben-style}
    Q_\perp = \frac{v^3_{\rm th,i}}{\rho_{\rm i}} \left[ c_1 \epsilon^3_{\rm i} \exp\left( -\frac{c_2}{\epsilon_{\rm i}}\right) \right]
\end{equation}
with $c_1 \sim 1 $ (typically $0.75$) and $c_2 \approx 0.1$--$0.3$ (typically $0.34$ or ${\simeq}0.2$). (Recall that the value of $c_\ast$ that best fits our simulation results is ${\approx} 0.05$--$0.1$.) Average values of $\epsilon_{\rm i}$ inferred between ${\sim}0.3~{\rm au}$ and ${\sim}1~{\rm au}$ from the Sun were in the range of ${\approx}0.03$--$0.05$.

The results of our paper suggest that the following refinements to this procedure may improve its accuracy. First, it is not necessarily the case that the fluctuations on ion gyroscales are accurately described by the Alfv\'{e}nic relation \eqref{eqn:AWKAW}. Indeed, the argument in \S\ref{subsec:theory_OhmLawArguments} is that the gyroscale potential fluctuations may be better inferred at low beta using $q_{\rm i}\delta\Phi_{\rho_{\rm i}}/m_{\rm i} v^2_{\rm th,i} \sim \beta^{-1}_{\perp} (\delta B_{\rho_{\rm i}}/B_0)$, rather than ${\sim}\delta u_{\rm i,\rho_{\rm i}}/v_{\rm th\perp,i} = \sigma \beta^{-1/2}_{\perp\rm i} (\delta B_{\rho_{\rm i}}/B_0)$. [Recall the definition $\beta_\perp = (1+\tau_\perp)\beta_{\perp\rm i}$.] While it is true that there are combinations of $\tau_\perp$ and $\beta_{\perp\rm i}$ for which these two formulae return comparable inferred potential fluctuations, the interpretative difference is notable -- at very low values of $\beta_\perp$, the electrostatic potential with which particles interact on their gyroscale has little to do with fluctuations in the ion flow velocity. When in doubt, a generalized Ohm's law that accounts for sub-$d_{\rm i}$ contributions to the electrostatic potential, such as Equation \eqref{eq:generalized_Ohm}, should be used.

To give concrete numbers, the rms fluctuation levels centered about the ion thermal Larmor scale in our $\beta_{\rm i0}=1/9$ simulation (calculated as in Equation \ref{eqn:Bcentered}) are $\delta B_{\rho_{\rm i}}/B_0 \simeq 0.042$ and $\delta u_{\rm i,\rho_{\rm i}}/v_{\rm A0} \simeq 0.024$; in our $\beta_{\rm i0}=0.3$ simulation, they are $\delta B_{\rho_{\rm i}}/B_0 \simeq 0.043$ and $\delta u_{\rm i,\rho_{\rm i}}/v_{\rm A0} \simeq 0.021$. Neither of these sets of values satisfy Equation \eqref{eqn:AWKAW} when $\sigma=1.19$, and both suggest $\sigma < 1$. In this context, it is worth noting that these ion-Larmor-scale magnetic-field fluctuation amplitudes are typical of (if just slightly larger than) those in the low-beta solar wind: \citet{Bourouaine13} used {\em Helios} data to report $\delta B_{\rho_{\rm i}}/B_0 \approx 0.03$ at ${\approx}0.3~{\rm au}$, while \citet{MartinovicAPJS2020} used {\em Parker Solar Probe} data to find strong evidence in the ion distribution function for stochastic heating at ${\approx}0.2~{\rm au}$ when $\delta B_{\rho_{\rm i}}/B_0 \simeq 0.049$ (see their figure 5(a)). Both authors used the relation \eqref{eqn:epsilon} with $\sigma=1.19$ to compute $\epsilon_{\rm i}$, reporting values in the range ${\approx}0.04$--$0.08$ when $\beta_{\rm i} \approx 0.3$--$0.5$. The stochasticity parameter in our $\beta_{\rm i0}=1/9$ run, based on rms potential fluctuations centered about $\rho_{\rm i}$, is notably larger at $\xi^{\rm (rms)}\simeq 0.146$; accounting for intermittency raises its value to $\xi^{\rm (eff)}\simeq 0.173$. In our $\beta_{\rm i0}=0.3$ run, we measured $\xi^{\rm (rms)} \simeq 0.085$ and $\xi^{\rm (eff)} \simeq 0.10$. Whether the difference between the observationally inferred $\epsilon_{\rm i}$ and the values of $\xi$ we obtained from the potential fluctuations in our simulations is primarily because Equation \eqref{eqn:AWKAW} is an inaccurate proxy for electrostatic potential fluctuations at low $\beta$, or because intermittency effects must be taken into account, or perhaps because our simulations could benefit from slightly larger scale separation, awaits more data (both actual and numerical) and further scrutiny. Given the exponential sensitivity of $Q_\perp$ in Equation \eqref{eq:Qprp_Ben-style} to $\epsilon_{\rm i}$, obtaining an accurate value of $c_2$ relies on an accurate definition of the stochasticity parameter.


\section{Conclusions}\label{sec:conclusions}

We have derived a generalization of the theory of stochastic ion heating originally presented in \citet{ChandranAPJ2010}, adapted to the case in which electric-field fluctuations can be described by a generalized Ohm's law that includes Hall and thermo-electric effects. 
We argued that these non-ideal terms provide the dominant contribution to the stochastic heating of ions at sub-$d_{\rm i}$ scales, which are the relevant scales at which stochastic heating operates in low-$\beta$ turbulence (i.e., when $\rho_{\rm i}\ll d_{\rm i}$). 
By keeping a fully scale-dependent approach, both in configuration space and in velocity space, we have derived the perpendicular-heating rate $Q_\perp$ and perpendicular-energy diffusion coefficient $D_{\perp\perp}^E$ as functions of the perpendicular ion velocity $w_\perp$ and the perpendicular plasma beta $\beta_\perp$, adopting certain well-established properties of inertial- and dispersion-range turbulent fluctuations.

The predictions of this theory were then tested using 3D hybrid-kinetic PIC simulations of continuously driven Alfv\'enic turbulence at low $\beta$, namely, the $\beta_{\rm i0}=0.3$ simulation presented by \citet{ArzamasskiyAPJ2019} and a newly performed $\beta_{\rm i0}=1/9$ simulation. In these simulations, parallel heating of ions is primarily associated with Landau/Barnes damping of turbulent fluctuations, and is always sub-dominant with respect to its perpendicular counterpart, $Q_{\|,{\rm i}}\ll Q_{\perp,{\rm i}}$. Two perpendicular-heating mechanisms are shown to operate simultaneously on ions and to provide most of their heating: {\em ion-cyclotron} and {\em stochastic} heating.
While ion-cyclotron dominates over stochastic heating at $\beta_{\rm i0}=0.3$ ($Q_{\perp,{\rm i}}^{\rm cycl}/Q_{\perp,{\rm i}}^{\rm tot}\gtrsim75$\% and $Q_{\perp,{\rm i}}^{\rm stoch}/Q_{\perp,{\rm i}}^{\rm tot}\lesssim25$\%), in the $\beta_{\rm i0}=1/9$ simulation these two mechanisms contribute roughly equally to the perpendicular heating of ions ($Q_{\perp,{\rm i}}^{\rm stoch}/Q_{\perp,{\rm i}}^{\rm tot}\approx Q_{\perp,{\rm i}}^{\rm cycl}/Q_{\perp,{\rm i}}^{\rm tot}\approx50$\%).
As far as stochastic ion heating is concerned, the theoretical predictions derived in this work describe reasonably well the associated features emerging from the simulations and characterized by various heating diagnostics, both in perpendicular-velocity and in perpendicular-wavevector spaces. These diagnostics also emphasize the important role of non-MHD contributions to the electrostatic potential in stochastically heating the ions at low $\beta$, and demonstrate that intermittency in the turbulence enhances this heating. Finally, the fraction of injected energy that is channeled into total ion heating strongly depends on the plasma $\beta$, passing from being $Q_{\rm i}^{\rm tot}/\varepsilon_{\rm AW}\approx75$\% at $\beta_{\rm i}=0.3$ to $Q_{\rm i}^{\rm tot}/\varepsilon_{\rm AW}\approx40$\% at $\beta_{\rm i}\approx 0.1$.

Our work has three main implications for the interpretation of spacecraft data in the context of stochastic heating. First, we have provided a number of phase-space diagnostics that one may use to supplement the presently employed technique of inferring stochastic heating in the solar wind via correlations between the amplitudes of ion-Larmor-scale magnetic fluctuations and plasma heating. These diagnostics supplement concurrent work on field-particle correlations by \citet{KleinHowesApJL2016}, \citet{HowesJPP2017}, and others, which show great promise in their ability to distinguish between various particle-energization mechanisms and their contributions to the heating of the solar wind. Second, the precise way in which spacecraft-measured, ion-Larmor-scale magnetic-field fluctuations are translated into electric potential fluctuations to calculate stochastic heating deserves careful re-examination, especially at $\beta$ values small enough that $d_{\rm i}\gg\rho_{\rm th,i}$. In particular, we advocate for the use of a generalized Ohm's law that accounts for the (sometimes dominant) contributions from the Hall and thermo-electric effects to the electric potential. We find that the implied stochasticity parameter $\xi_{\rm th} = q_{\rm i}\delta\Phi_{\rho_{\rm i}}/m_{\rm i}v_{\rm th,i}^2$ obtained from the full potential fluctuations is generally larger than that implied by Equation \eqref{eqn:epsilon}, particularly when intermittency effects are taken into account. Third, our simulation results suggest a link between preferential perpendicular heating, magnetic spectra that exhibit sub-ion-Larmor steepening, and perpendicular distribution functions with flattened cores -- a link which, if due to stochastic heating, should be pronounced when the amplitude of ion-Larmor-scale magnetic fluctuations is relatively large ({\em viz.}, $\xi \gtrsim 0.1$). 

With the gradual decrease in the perihelion of {\em Parker Solar Probe} \citep{Fox2016SSRv}, and the increasing level of turbulent activity towards the Alfv\'{e}n point \citep{TuMarsch1995SSRv,ChandranAPJ2011,BrunoCarboneLRSP2013,ChenAPJS2020}, the importance of understanding the phase-space signatures of stochastic heating will only become greater. It is our hope that the predictions and diagnostics presented here will help to sharpen this understanding and facilitate a more robust analysis of current and future spacecraft data.

\acknowledgments

It is a pleasure to thank Kristopher Klein, Mihailo Martinovi\'{c}, and Benjamin Chandran for extremely useful discussions on stochastic ion heating and the interpretation of spacecraft measurements, and Phil Isenberg, Bernie Vasquez, and the referee for constructive comments on the manuscript. We further acknowledge PRACE for awarding us access to the supercomputer Marconi, CINECA, Italy, where our $\beta_{\rm i0}=1/9$ simulation was performed, under grant n.~2017174107. This research was supported by NASA Grant No.~NNX16AK09G issued through the Heliophysics Supporting Research Program, by the Max-Planck/Princeton Center for Plasma Physics (NSF grant PHY-1804048), and by an Alfred P. Sloan Research Fellowship in Physics to M.W.K. L.A. gratefully acknowledges support from the Institute for Advanced Study.

\appendix

\section{A.~Alternative heating diagnostics}\label{app:sec:diffusion}

In this Appendix, we summarize the implementation of the heating diagnostic in \textsc{Pegasus}\texttt{++} and discuss its limitations. We begin by reviewing our definition of the perpendicular-energy diffusion coefficient. If the particle heating occurs through a diffusion-like process, the distribution function evolves according to 
\begin{equation}
    \frac{\partial f^{E}}{\partial t} = \frac{\partial}{\partial e_\perp} \left(D^E_{\perp\perp}\frac{\partial f^{E}}{\partial e_\perp} \right),\label{eq:app_diffusion}
\end{equation}
where $f^{E}$ is the perpendicular-energy distribution function. Here we assume that energization is only perpendicular to the magnetic field, $e_\perp \doteq w_\perp^2/2$. The total energy of the distribution is $E \doteq \int{\rm d} e_\perp\, e_\perp f^{E}$. The total energization rate is thus
\begin{equation}
     Q_\perp \doteq \D{t}{E} = \D{t}{} \int\rmd e_\perp \, e_\perp f^{E} = \int\rmd e_\perp \, e_\perp \pD{t}{f^E} = \int\rmd e_\perp \,  e_\perp \pD{e_\perp}{} \left( D^{E}_{\perp\perp} \pD{e_\perp}{f^E} \right) .
\end{equation}
In a collisionless plasma, this energization is provided by the work performed by the electric fields on the particle distribution,
\begin{equation}
    q_\perp(e_\perp) \doteq q_{\rm i} \bb{v}_\perp\bcdot\bb{E}_\perp.
\end{equation}
The perpendicular-energy distribution function of ions evolves according to a Vlasov equation of form 
\begin{equation}
    \frac{\partial f^E}{\partial t} + \frac{\partial}{\partial e_\perp} \biggl(q_\perp(e_\perp) f^E\biggr) = 0;
\end{equation}
this equation neglects terms related to advection and to exchange between parallel and perpendicular energies (e.g., pitch-angle scattering), but retains some basic properties of the kinetic equation, such as the conservation of particle number. The equation for the diffusion coefficient is then simply
\begin{equation}
   \frac{\partial}{\partial e_\perp} \left(D^E_{\perp\perp}\frac{\partial f^{E}}{\partial e_\perp} \right) + \frac{\partial}{\partial e_\perp} \biggl(q_\perp(e_\perp) f^E\biggr) = 0 \qquad \Longrightarrow \qquad D_{\perp\perp}^E = -\frac{q_\perp(e_\perp) f^E}{\partial f^E/\partial e_\perp},
\end{equation}
which could also be written as 
\begin{equation}
    D^{E}_{\perp\perp} = - \frac{\partial Q_\perp/\partial e_\perp}{\partial f^E/\partial e_\perp},\label{eq:diff_num}
\end{equation}
where we define the differential heating rate as 
\begin{equation}
    \frac{\partial Q}{\partial \bb{v}} \doteq q_{\rm i}\bb{v}\bcdot\bb{E} f. \label{eq:dQdv_def}
\end{equation}
We use the definition (\ref{eq:diff_num}) of the energy diffusion coefficient throughout this paper (as well as in \citealt{ArzamasskiyAPJ2019}) to describe the velocity-space dependence of ion heating. At large scales, energization (\ref{eq:dQdv_def}) is dominated by the conversion between bulk-kinetic and magnetic energies related to Alfv\'enic motions. In order to remove this non-dissipative process, we use $\bb{w} = \bb{v} - \bb{u}$ instead of $\bb{v}$ for the heating diagnostics and perform a long-time average.

\begin{figure}
    \centering
    \includegraphics[width=\textwidth]{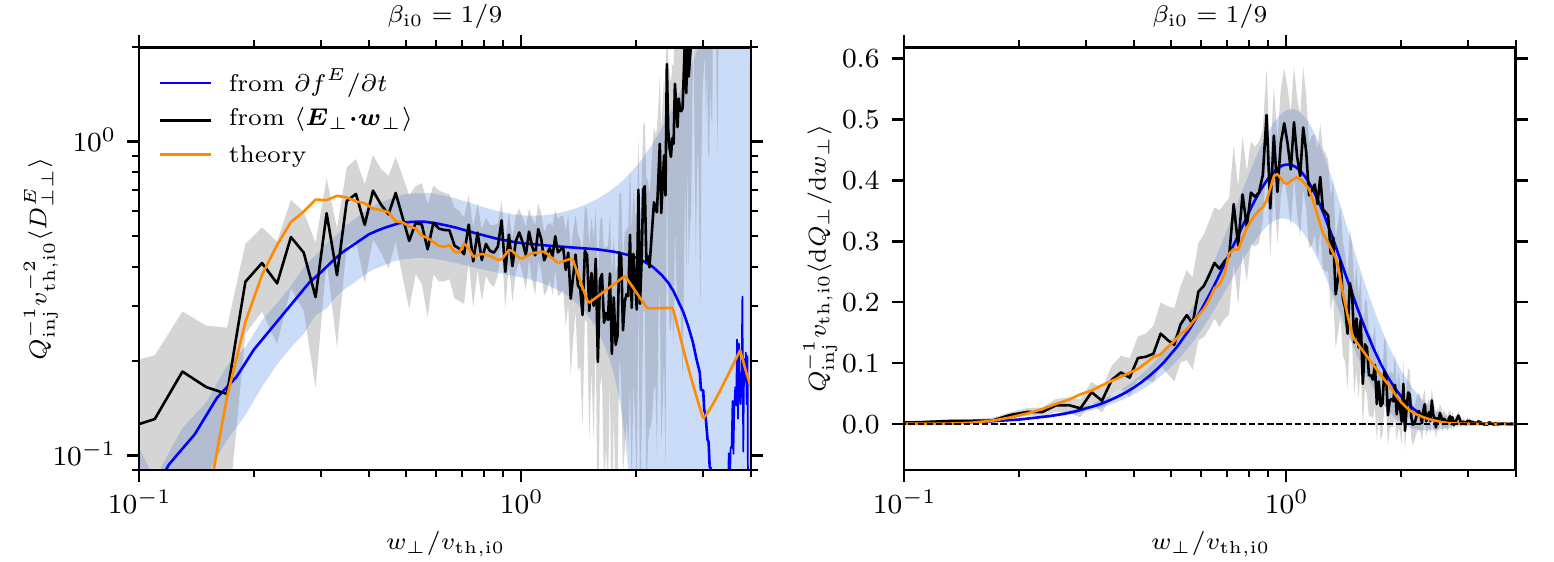}
    \caption{Comparison between two methods for computing the perpendicular energy diffusion coefficient. The blue line is obtained from the evolution of energy distribution function following the method of \cite{VasquezAPJ2020}. The black line is computed using the $\bb{E}\bcdot\bb{w}$ diagnostic used throughout this paper and in \cite{ArzamasskiyAPJ2019}. The orange line represents the theoretical prediction for stochastic heating based on the electrostatic potential fluctuations (Equation~\ref{eq:Dperp_Phi_general-b}). (Left) Energy diffusion coefficient. (Right) Velocity-space dependence of ion energization. The shaded regions represent the time-variability of plotted quantities (computed as a standard deviation).}
    \label{fig:diffusion_comparison}
\end{figure}

Recently, \cite{VasquezAPJ2020} argued that the energy diffusion coefficient should be defined differently. They argued that differential energization should be equal to 
\begin{equation}
    \pD{e_\perp}{Q_\perp} = \pD{t}{e_\perp f^E}. \label{eq:dQde_vasq}
\end{equation}
Using this definition, they arrived at a more complicated equation for the diffusion coefficient:
\begin{equation}
    \pD{e_\perp}{Q_\perp} =  - D_{\perp \perp}^{E} \pD{e_\perp}{f^E} + \pD{e_\perp}{} \left(e_\perp D_{\perp\perp}^{E} \pD{e_\perp}{f^E} \right) , \label{eq:diff_vasq}
\end{equation}
which has an additional term relative to (\ref{eq:diff_num}). The difference comes from the definition of $\partial Q_\perp/\partial e_\perp$. This quantity is not well-defined: if one adds any derivative of form $\partial F/\partial e_\perp$ to $\partial Q_\perp/\partial e_\perp$ with $F|_0^{\infty} = 0$, the total heating rate $Q_\perp$ remains unchanged. Indeed, Equation (\ref{eq:diff_vasq}) differs from (\ref{eq:diff_num}) by such a term. If one were to use the definition (\ref{eq:dQdv_def}), then the appropriate definition of the diffusion coefficient is (\ref{eq:diff_num}).

These two methods for calculating $D_{\perp \perp}^{E}$ require very different numerical implementations. In order to use the method of \cite{VasquezAPJ2020}, one only needs to measure the distribution function at different moments in time, and then solve Equation (\ref{eq:dQde_vasq}). In contrast, to use Equation (\ref{eq:diff_num}) one needs both $\partial Q_\perp/\partial e_\perp$ and $f^E$, but the equation for $D_{\perp \perp}^{E}$ becomes much easier to solve. 

Figure \ref{fig:diffusion_comparison} shows the comparison of energy diffusion coefficients (left) and energization (right) as functions of velocity space computed using the evolution of the energy distribution function (blue) and using our $\bb{E}\bcdot \bb{v}$ diagnostic (black). The blue curve is normalized to the total heating rate while the black curve has slightly different normalization so that diffusion coefficient has the same magnitude in the $w_\perp \ll v_{\rm th,i0}$ part of the plot. We conclude that both methods produce very similar results in the  $w_\perp \lesssim v_{\rm th,i0}$ part of velocity space, where our stochastic-heating theory is expected to work best.

\section{B.~Exact calculation of $Q_\perp$ with exponential correction and its limits}\label{app:sec:exact_calculation_Qperp}
    
In this Appendix, we use Equation \eqref{eq:Dperp_Phi_general_xi_EXPcorr} for the diffusion coefficient including the exponential correction to derive formulae for the implied perpendicular heating. We begin with Equation \eqref{eq:Dperp_Phi_general_xi_EXPcorr} written in terms of the potential fluctuations,
\begin{equation}\label{app:eq:Dperp_Phi_general_xi_EXPcorr}
   D_{\perp\perp}^E(w_\perp) \sim \Omega_{\rm i}\,m_{\rm i}^{2} v_{\rm th,i}^{4}\left[\left(\frac{w_\perp}{v_{\rm th,i}}\right)^{-2}\frac{q_{\rm i}^3|\delta\Phi_w|^3}{m_{\rm i}^3v_{\rm th,i}^6}\right]\exp\left[-c_*\left(\frac{w_\perp}{v_{\rm th,i}}\right)^2\frac{m_{\rm i}v_{\rm th,i}^2}{q_{\rm i}\delta\Phi_{w}}\right],
\end{equation}
which is then substituted into the perpendicular-heating integral, 
\begin{equation}\label{app:eq:Qperp-integral}
   Q_\perp = -\int_0^\infty \rmd w_\perp \,D_{\perp\perp}^E \pD{w_\perp}{f^E}.
\end{equation}
We then evaluate the result in the two limits considered in \S\ref{subsec:theory_AW-KAW_scaling}, namely, $\beta_\perp\gtrsim 1$, for which the inductive electric field dominates the ion-gyroscale electrostatic potential, and $\beta_\perp\ll 1$, for which the ion-gyroscale fluctuations are predominantly sub-$d_{\rm i}$ KAWs. 

\subsection{Stochastic heating with exponential correction in $\beta\gtrsim{1}$ AW turbulence}

In this limit, the electrostatic potential evaluated at perpendicular velocity $w_\perp \sim v_{\rm th,i}(\lambda/\rho_{\rm th,i})$ is given by
\begin{equation}\label{app:eq:Phi-w_AWlimit}
   \delta\Phi_w
   \sim
   \rho_{\rm th,i}\left(\frac{w_\perp}{v_{\rm th,i}}\right)\frac{\delta u_{{\rm \perp i},w}}{c}\,B_0
   \sim
   \rho_{\rm th,i}B_0\frac{v_{\rm th,i}}{c}\left(\frac{\varepsilon_{\rm AW}}{\Omega_{\rm i}v_{\rm A}^2}\right)^{1/3}\beta_{\rm \perp i}^{-1/3}\left(\frac{w_\perp}{v_{\rm th,i}}\right)^{4/3},
\end{equation}
where we have used Equation \eqref{eq:dUperp_scaling_AW} to rewrite $\delta\Phi_\lambda\to\delta\Phi_w$. The corresponding diffusion coefficient is then
\begin{equation}\label{app:eq:Dperp_Phi_EXPcorr_AWlimit}
   D_{\perp\perp}^{\rm(AW)}(w_\perp)
   \sim
   \varepsilon_{\rm AW}\, m_{\rm i}^2 v_{\rm th,i}^2\left(\frac{w_\perp}{v_{\rm th,i}}\right)^2\exp\left[-\mu_*^{\rm(AW)}\left(\frac{w_\perp}{v_{\rm th,i}}\right)^{2/3}\right],
\end{equation}
where 
\begin{equation}\label{app:eq:mu-star_AWlimit}
   \mu_*^{\rm(AW)}\,
   \doteq\,
   c_*\beta_{\rm \perp i}^{1/3}\left(\frac{\Omega_{\rm i}v_{\rm A}^2}{\varepsilon_{\rm AW}}\right)^{1/3}\,
   =\,
   c_*\beta_{\rm \perp i}^{1/2}\left(\frac{L}{\rho_{\rm th,i}}\right)^{1/3}.
\end{equation}
In the final step above, we have used $\varepsilon_{\rm AW}=v_{\rm A}^3/L$ to relate $\mu_*$ to the separation of scales in the system. 
Using $f^E(w_\perp)=\exp(-w_\perp^2/v_{\rm th,i}^2)/(m_{\rm i}v_{\rm th,i}^2)$ and rewriting $\exp[-\mu_*^{\rm(AW)}(w_\perp/v_{\rm th,i})^{2/3}]$ using the definitions of the exponential and Gamma functions, {\em viz.}, $\exp(x)=\sum_{n=0}^\infty x^n / n!$ and $n! = \Gamma(n+1)$, respectively, we can perform the integral in Equation \eqref{app:eq:Qperp-integral} to determine the heating rate per unit mass of stochastic heating off of AW fluctuations, $Q_\perp^{\rm(AW)}$:
\begin{align}\label{app:eq:Qperp-integral_AWlimit}
   \frac{Q_\perp^{\rm(AW)}}{\varepsilon_{\rm AW}} &\sim 2\int_0^\infty \frac{\rmd w_\perp}{v_{\rm th,i}} \, \left(\frac{w_\perp}{v_{\rm th,i}}\right)^{3} \exp\left[-\left(\frac{w_\perp}{v_{\rm th,i}}\right)^2-\mu_*^{\rm(AW)}\left(\frac{w_\perp}{v_{\rm th,i}}\right)^{2/3}\right]\nonumber\\
   \mbox{} &= \sum_{n=0}^{\infty} \frac{\left(-\mu_*^{\rm(AW)}\right)^n}{\Gamma(n+1)} \, 2\int_0^\infty\rmd x\, x^{3+2n/3} \,{\rm e}^{-x^2} \qquad ({\rm with}~x\doteq w_\perp/v_{\rm th,i})\nonumber\\
   \mbox{} &= \sum_{n=0}^{\infty} \frac{\Gamma(2+n/3)}{\Gamma(n+1)}\left(-\mu_*^{\rm(AW)}\right)^n \nonumber\\
   & \nonumber\\
  \Longrightarrow \frac{Q_\perp^{\rm(AW)}}{\varepsilon_{\rm AW}} &=  \Lambda_{\rm AW} \sum_{n=0}^{\infty}\frac{\Gamma(2+n/3)}{\Gamma(n+1)}\left(-\mu_*^{\rm(AW)}\right)^n .
\end{align}
As in Equation \eqref{eq:Qperp_Phi_scaling_AW}, $\Lambda_{\rm AW}$ is a constant independent of $\beta_{\perp\rm i}$ and $\tau_\perp$ that takes into account the various coefficients neglected in our scaling arguments. Note that $\Gamma(2+n/3)/\Gamma(n+1)$ is a function that quickly decreases for $n>2$; for $n=0,1,2$, its values are $1$, ${\simeq}1.19$, ${\simeq}0.75$. Although the result of the integral in equation \eqref{app:eq:Qperp-integral_AWlimit} is exact, it is worth specifying its approximations in two regimes, $\mu_*^{\rm(AW)}\lesssim1$ and $\mu_*^{\rm(AW)}\gg1$.

\paragraph{\bf $\boldsymbol{\mu_*^{\rm(AW)}\lesssim1}$ regime.}This is the regime in which stochastic heating is most efficient. The condition $\mu_*^{\rm(AW)}\lesssim1$ is met as long as
\begin{equation}\label{app:eq:mu-star_AWlimit_less-than-one}
   \varepsilon_{\rm AW}\gtrsim \varepsilon_{\rm crit}^{\rm(AW)}\doteq c_*^3 \beta_{\rm \perp i}\Omega_{\rm i}v_{\rm A}^2
   \qquad
   {\rm or}\qquad
   \frac{\rho_{\rm th,i}}{L} \gtrsim
   \chi_{\rm crit}^{\rm(AW)}\doteq c_*^3\beta_{\rm \perp i}^{3/2}.
\end{equation}
Therefore, if the system is such that the energy injected in the Alfv\'enic cascade exceeds a certain critical value $\varepsilon_{\rm crit}^{\rm(AW)}$ (or, equivalently, if the scale separation $\rho_{\rm th,i}/L$ remains above a critical value $\chi_{\rm crit}^{\rm(AW)}$), then the dominant contributions to \eqref{app:eq:Qperp-integral_AWlimit} are the $n=0,1$ terms. As a result, 
\begin{equation}\label{app:eq:Qperp-integral_AWlimit_mu-less-than-one}
   \frac{Q_\perp^{\rm(AW)}}{\varepsilon_{\rm AW}} 
   \approx 
   \Lambda_{\rm AW}\left[1 - \Gamma(7/3)\,c_*\beta_{\rm \perp i}^{1/2}\left(\frac{L}{\rho_{\rm th,i}}\right)^{1/3}\right].
\end{equation}
The second term in brackets is a small correction to the expression \eqref{eq:Qperp_Phi_scaling_AW} obtained by neglecting the exponential suppression.

\paragraph{\bf $\boldsymbol{\mu_*^{\rm(AW)}\gg1}$ regime.}This is the regime in which stochastic heating is strongly suppressed for most of the ion population by the quasi-conservation of their magnetic moment. This regime holds, for instance, when the separation between the injection scale and the ion-Larmor scale in a system is significantly larger than the critical value derived above, i.e., when 
\begin{equation}\label{app:eq:mu-star_AWlimit_much-larger-than-one}
   \frac{\rho_{\rm th,i}}{L}\,\ll\,
   \chi_{\rm crit}^{\rm(AW)}\doteq\,c_*^3\beta_{\rm \perp i}^{3/2}.
\end{equation}
To obtain $Q_\perp^{\rm(AW)}$ in this limit, it is easier to make some approximations before performing the integral. Namely, when the exponential suppression factor is important, we may safely neglect the $\exp(-w_\perp^2/v_{\rm th,i}^2)$ term in the integral \eqref{app:eq:Qperp-integral_AWlimit}. In this case, the resulting heating is 
\begin{equation}\label{app:eq:Qperp-integral_AWlimit_mu-much-larger-than-one}
   \frac{Q_\perp^{\rm(AW)}}{\varepsilon_{\rm AW}}
   \approx
   \frac{\Lambda_{\rm AW}}{\left(\mu_*^{\rm(AW)}\right)^6}
   =
   \Lambda_{\rm AW} \,c_*^{-6}\beta_{\rm \perp i}^{-3}\left(\frac{\rho_{\rm th,i}}{L}\right)^2.
\end{equation}

\subsection{Stochastic heating with exponential correction in low-$\beta$ KAW turbulence}

In this limit, the electrostatic potential fluctuations may be approximated by
\begin{equation}\label{app:eq:Phi-w_KAWlimit}
   \delta\Phi_w
   \approx
   d_{\rm i}\left(1+\tau_\perp\right)^{-1}\frac{\delta B_{\|,w}}{B_0}\,\frac{v_{\rm A}}{c}B_0 \sim
   \rho_{\rm th,i}B_0\frac{v_{\rm th,i}}{c}\left(\frac{\varepsilon_{\rm KAW}}{\Omega_{\rm i}v_{\rm A}^2}\right)^{1/3}\frac{(1+\tau_\perp)^{-2/3}\beta_{\rm \perp i}^{-1/3}}{(2+\beta_\perp)^{1/3}}\left(\frac{w_\perp}{v_{\rm th,i}}\right)^{(3+\alpha)/6},
\end{equation}
and thus
\begin{equation}\label{app:eq:Dperp_Phi_EXPcorr_KAWlimit}
   D_{\perp\perp}^{\rm(KAW)}(w_\perp)
   \sim
   \varepsilon_{\rm KAW}m_{\rm i}^2 v_{\rm th,i}^2\frac{(1+\tau_\perp)^{-2}}{(2+\beta_\perp)}\left(\frac{w_\perp}{v_{\rm th,i}}\right)^{(\alpha-1)/2}\exp\left[-\mu_*^{\rm(KAW)}\left(\frac{w_\perp}{v_{\rm th,i}}\right)^{(9-\alpha)/6}\right],
\end{equation}
where now the parameter $\mu_*$ is defined by
\begin{equation}\label{app:eq:mu-star_KAWlimit}
   \mu_*^{\rm(KAW)}
   \doteq
   c_*\beta_{\rm \perp i}^{1/3}(1+\tau_\perp)^{2/3}(2+\beta_\perp)^{1/3}\left(\frac{\Omega_{\rm i}v_{\rm A}^2}{\varepsilon_{\rm KAW}}\right)^{1/3}
   =
   c_*\beta_{\rm \perp i}^{1/2}(1+\tau_\perp)^{2/3}(2+\beta_\perp)^{1/3}\left(\frac{\varepsilon_{\rm AW}}{\varepsilon_{\rm KAW}}\right)^{1/3}\left(\frac{L}{\rho_{\rm th,i}}\right)^{1/3}.
\end{equation}
We remind the reader that $1\leq\alpha\leq3$ is the parameter taking into account different models for the spectral anisotropy of the cascading KAW fluctuations (see Equation \eqref{eq:KAWanisotropy_alpha} in \S\ref{subsec:theory_AW-KAW_scaling}).
Performing the $w_\perp$-integral of $-D_{\perp\perp}^{\rm(KAW)}(\partial f^E/\partial w_\perp)$ and proceeding as in the AW case, we find that the heating rate per unit mass of stochastic heating off of KAW fluctuations satisfies
\begin{align}\label{app:eq:Qperp-integral_KAWlimit}
   \frac{Q_\perp^{\rm(KAW)}}{\varepsilon_{\rm KAW}} &\sim 2\int_0^\infty \frac{\rmd w_\perp}{v_{\rm th,i}} \, \left(\frac{w_\perp}{v_{\rm th,i}}\right)^{(\alpha+1)/2}\exp\left[-\left(\frac{w_\perp}{v_{\rm th,i}}\right)^2-\mu_*^{\rm(KAW)}\left(\frac{w_\perp}{v_{\rm th,i}}\right)^{(9-\alpha)/6}\right] \nonumber\\
   \mbox{} &=
   \frac{(1+\tau_\perp)^{-2}}{(2+\beta_\perp)}\sum_{n=0}^{\infty} \frac{\left(-\mu_*^{\rm(KAW)}\right)^n}{\Gamma(n+1)} 2\int_0^\infty \rmd x\, x^{(\alpha+1)/2+n(9-\alpha)/6} \,{\rm e}^{-x^2} \nonumber\\
  \mbox{} &= \frac{(1+\tau_\perp)^{-2}}{(2+\beta_\perp)}\sum_{n=0}^{\infty} \frac{\Gamma\left(\frac{3+\alpha}{4}+n\frac{9-\alpha}{12}\right)}{\Gamma(n+1)}\left(-\mu_*^{\rm(KAW)}\right)^n\nonumber\\
  & \nonumber\\
  \Longrightarrow\frac{Q_\perp^{\rm(KAW)}}{\varepsilon_{\rm KAW}} &=
  \Lambda_{\rm KAW}
  \frac{(1+\tau_\perp)^{-2}}{(2+\beta_\perp)}\,\sum_{n=0}^{\infty}\frac{\Gamma\left(\frac{3+\alpha}{4}+n\frac{9-\alpha}{12}\right)}{\Gamma\left(\frac{3+\alpha}{4}\right)\Gamma(n+1)}\left(-\mu_*^{\rm(KAW)}\right)^n .
\end{align}
As in Equation \eqref{eq:Qperp_Phi_scaling_KAW}, $\Lambda_{\rm KAW}$ is a constant independent of $\beta_{\perp\rm i}$ and $\tau_\perp$ that takes into account the various coefficients neglected in our scaling arguments (note that a factor $\Gamma\left(\frac{3+\alpha}{4}\right)$ has been introduced in the denominator within the sum, so that the $n=0$ term exactly matches the expression in \eqref{eq:Qperp_Phi_scaling_KAW}; this is also absorbed within the constant $\Lambda_{\rm KAW}$). Once again, Equation \eqref{app:eq:Qperp-integral_KAWlimit} is exact, but it is instructive to derive explicit analytical expressions for $Q_\perp^{\rm(KAW)}$ in the two interesting limits, $\mu_*^{\rm(KAW)}\lesssim1$ and $\mu_*^{\rm(KAW)}\gg1$.

\paragraph{\bf $\boldsymbol{\mu_*^{\rm(AW)}\lesssim1}$ regime.}This is the case for which the quasi-conservation of the magnetic moment does not effectively hold, making the stochastic heating of ions more effective. Such regime occurs if the energy cascading as KAW fluctuations exceeds a critical energy cascade rate $\varepsilon_{\rm crit}^{\rm(KAW)}$ given by
\begin{equation}\label{app:eq:mu-star_KAWlimit_less-than-one_A}
   \varepsilon_{\rm KAW}
   \gtrsim
   \varepsilon_{\rm crit}^{\rm(KAW)}
   \doteq
   c_*^3\, \beta_{\rm \perp i}(1+\tau_\perp)^{2}(2+\beta_\perp)\Omega_{\rm i}v_{\rm A}^2,
\end{equation}
or, in other words, if the scale separation $\rho_{\rm th,i}/L$ in the system remains above a critical value $\chi_{\rm crit}^{\rm(KAW)}$ given by
\begin{equation}\label{app:eq:mu-star_KAWlimit_less-than-one_B}
   \frac{\rho_{\rm th,i}}{L}
   \gtrsim
   \chi_{\rm crit}^{\rm(KAW)}
   \doteq
   \left(\frac{\varepsilon_{\rm AW}}{\varepsilon_{\rm KAW}}\right)\,c_*^3\, \beta_{\rm \perp i}^{3/2}(1+\tau_\perp)^{2}(2+\beta_\perp).
\end{equation}
Retaining only the $n=0,1$ terms in Equation \eqref{app:eq:Qperp-integral_KAWlimit}, we may approximate the heating in this limit as  
\begin{equation}\label{app:eq:Qperp-integral_KAWlimit_mu-less-than-one}
   \frac{Q_\perp^{\rm(KAW)}}{\varepsilon_{\rm AW}}
   \approx
   \Lambda_{\rm KAW}\left(\frac{\varepsilon_{\rm KAW}}{\varepsilon_{\rm AW}}\right)\frac{(1+\tau_\perp)^{-2}}{(2+\beta_\perp)}\left[1-c_*\,\frac{\Gamma\left(\frac{9+\alpha}{6}\right)}{\Gamma\left(\frac{3+\alpha}{4}\right)}\, \beta_{\rm \perp i}^{1/2}(1+\tau_\perp)^{2/3}(2+\beta_\perp)^{1/3}\left(\frac{\varepsilon_{\rm AW}}{\varepsilon_{\rm KAW}}\right)^{1/3}\left(\frac{L}{\rho_{\rm th,i}}\right)^{1/3}\right] .
\end{equation}
The second term in brackets is a small correction to the expression \eqref{eq:Qperp_Phi_scaling_KAW} obtained by neglecting the exponential suppression.

\paragraph{\bf $\boldsymbol{\mu_*^{\rm(AW)}\gg1}$ regime.}Here we consider once more the regime in which the ions' magnetic moments are quasi-conserved, i.e., the regime of asymptotically weak stochastic heating from KAW fluctuations. 
Proceeding as in the AW case, we neglect the $\exp(-w_\perp^2/v_{\rm th,i}^2)$ term with respect to the suppression $\exp[-\mu_*^{\rm(KAW)}(w_\perp/v_{\rm th,i})^{(9-\alpha)/6}]$ in the integral leading to Equation \eqref{app:eq:Qperp-integral_KAWlimit} and  obtain the following approximate expression:
\begin{align}\label{app:eq:Qperp-integral_KAWlimit_mu-much-larger-than-one}
   \frac{Q_\perp^{\rm(KAW)}}{\varepsilon_{\rm AW}} &\approx 
   \Lambda_{\rm KAW}\left(\frac{\varepsilon_{\rm AW}}{\varepsilon_{\rm KAW}}\right)\frac{(1+\tau_\perp)^{-2}}{(2+\beta_\perp)}\left(\mu_*^{\rm(KAW)}\right)^{-3(3+\alpha)/(9-\alpha)}\nonumber\\
   \mbox{} &=
   \Lambda_{\rm KAW} \left[ \left(\frac{\varepsilon_{\rm AW}}{\varepsilon_{\rm KAW}}\right) (1+\tau_\perp)^{-2} (2+\beta_\perp)^{-1} \right]^{12/(9-\alpha)} \left[ c^3_* \,\beta^{-3/2}_{\perp\rm i} \left(\frac{\rho_{\rm th,i}}{L}\right)\right]^{(3+\alpha)/(9-\alpha)} .
\end{align}
%

\vspace{3ex}
\bibliographystyle{apj}

\end{document}